\documentclass[aip, reprint, groupedaddress]{revtex4-2}

\usepackage{bm}
\usepackage[ruled,vlined]{algorithm2e}
\usepackage{graphicx}
\usepackage{amsmath,amssymb}
\usepackage{mathrsfs}
\usepackage{amsthm}
\usepackage{mathtools}
\usepackage{xcolor}
\usepackage{subcaption}
\usepackage{siunitx} 
\newcolumntype{T}{S[table-format=3.5]}
\usepackage{cellspace}%
\usepackage{float}

\DeclareMathOperator*{\argmin}{arg\,min}

\newcommand\REV[1]{{\color{black}#1}}

\makeatletter
\def\LT@LR@e{\LTleft\z@   \LTright\z@}%
\makeatother

\begin{document}



\title{Exploring the free-energy landscape of a rotating superfluid}


\author{Andrew Cleary}
\email[]{andrew.cleary@ed.ac.uk}
\affiliation{School of Mathematics, University of Edinburgh, Edinburgh, EH9 3FD, UK}

\author{Jacob Page}
\email[]{jacob.page@ed.ac.uk}
\affiliation{School of Mathematics, University of Edinburgh, Edinburgh, EH9 3FD, UK}


\date{\today}

\begin{abstract}
\noindent
The equilibrium state of a superfluid in a rotating cylindrical vessel is a vortex crystal -- an array of vortex lines which is stationary in the rotating frame. 
Experimental realisations of this behaviour typically show a sequence of transient states before the free-energy minimising configuration is reached. 
Motivated by these observations, we construct a new method for a systematic exploration of the free-energy landscape via gradient-based optimisation of a scalar loss function. 
Our approach is inspired by the pioneering numerical work of Campbell \& Ziff (\emph{Phys. Rev. B.} \textbf{20}, 1979), and makes use of automatic differentiation which crucially allows us to include entire solution trajectories in the loss. 
We first use the method to converge thousands of low-free-energy relative equilibria \REV{in the unbounded domain} for vortex numbers $10 \leq N \leq 30$, which reveals an extremely dense set of solutions. 
As part of this search, we discover new continuous families of relative equilibria which are often global minimisers of the free energy. 
These continuous families all consist of crystals arranged in a double-ring configuration, and we assess which state from the family is most likely to be observed experimentally by computing energy-minimising pathways from nearby local minima -- identifying a common entry point into the family. 
\REV{The continuous families become discrete sets of equal-energy solutions when the wall is introduced in the problem.}
Finally, we develop an approach to compute homoclinic orbits and use it to examine the dynamics in the vicinity of the minimising state by converging connections for low-energy saddles.
\end{abstract}


\maketitle

\begin{quotation}

A superfluid confined to a rotating disc undergoes a sequence of dissipative transitions through a series of metastable vortex crystals en route to the free-energy minimising state.
We develop a suite of new computational methods to map out the free-energy landscape around the global minimum, capable of searching for equilibria with specified properties, dynamical connecting orbits and non-dynamical energy-minimising pathways between states. 
As part of our search we discover previously unknown continuous families of double-ringed vortex crystals. 
\REV{Crystals within the families are generally global energy-minimisers, and the families exhibit a preferred symmetrical state through which these sequences of transitions enter. }
\end{quotation}

\section{Introduction \label{sec:introduction}}
Helium remains liquid at temperatures lower than any other substance, where it displays quantum mechanical behaviour on a macroscopic scale. 
Liquid ${}^4\text{He}$ undergoes a so-called $\lambda$-transition to a `superfluid’ phase as the temperature drops below 2.2K, at which point it exhibits a range of highly counter-intuitive properties including a complete absence of internal friction \cite{FEYNMAN195517}. 
The irrotational nature of a superfluid results in the formation of vortex lines in a rotating container, each with quantised circulation in units of $h / m$ (where $h$ is Planck's constant and $m$ is the mass of a ${}^4$He atom), rather than a solid body rotation, such that the rotation rate of the vortex crystal matches $\Omega$, the imposed angular velocity \cite{StatPhys,hess1967}. 
The vortex crystal is an exact equilibrium solution of the governing equations in a rotating frame, with the state observed at long time being a global minimiser of the free energy. 
Initial theoretical predictions of this phenomenon were verified experimentally \citep{yarmchuk1979}, with more recent experiments in other configurations (isomorphic to the two dimensional Euler equations) allowing for an increasingly detailed picture \citep{BELattice2001} where a relaxation onto the minimising state is preceded by the transient appearance of other (presumably unstable) vortex crystals \citep{Grzybowski2000}.

Campbell and Ziff \cite{campbell1979vortex, Campbell1978} performed an extensive numerical study to search for the free-energy minimising state of a collection of identical point vortices for various numbers of vortices, $1 \le N \le 30$ (and some special cases with larger $N$) \REV{in the unbounded domain}. 
In their approach, Campbell and Ziff converged vortex crystals (relative equilibria of the equations of motion) by initialising point vortices on a series of concentric circles and performing gradient descent on the free energy to iteratively update the vortex positions until a minimum was reached. 
\REV{This methodology complements} analytical approaches to find vortex crystals, which often rely on assuming certain symmetries or geometric constraints \cite{stieltjes1900, lewis1996, aref2005} and hence have often been restricted to fairly modest values of $N$. 
\REV{While many known equilibria are symmetric \citep{Aref2003}, more recent numerical work \citep{Aref1998,Newton2007} hints at the ubiquity of \emph{asymmetric} states, and}
recent reviews \cite{Newton2009, Newton2014} have highlighted the need for a systematic study of the low-energy equilibria that emerge as the number of vortices increase. 
We tackle this problem here by combining numerical methods typically employed in the dynamical systems approach to turbulence \citep{Viswanath2007}, and newer optimisation techniques that have emerged in the field of deep learning \citep{blondel2022efficient}.

The dynamical systems picture of turbulence envisions fluid motion as a trajectory in a high dimensional state space, pin-balling between simple invariant solutions \citep{Kawahara2012,Graham2021}. 
These ideas gained significant traction following the discovery of a myriad of unstable travelling waves on the laminar-turbulent boundary in a pipe \citep{Faisst2003,wedin_kerswell_2004} and with the discovery of an unstable periodic orbit embedded in the turbulent attractor of a minimal Couette flow \citep{Kawahara2001}. 
Finding these exact solutions relies on both (1) a sensible method for generating initial guesses for simple invariant sets and (2) a Newton-Raphson approach to converge them \citep{Viswanath2007,Chandler2013}. 
Point (1) here has been a significant constraint on applying these ideas at high Reynolds numbers, while point (2) restricts the search to (relative) equilibria and periodic orbits — the method is not appropriate for computing connecting orbits for example. 

Recent work in dynamical systems for fluids has benefited from the rapid advances in data driven methods and machine learning \citep{Pathak2018,Linot2020}. 
This has included new methods for generating plausible guesses for simple invariant solutions \citep{Page2020}, including using deep learning architectures \citep{Page2021}.
However, it is perhaps the development of automatic differentiation (AD) which offers the most compelling opportunities when combined with more traditional numerical simulation techniques.
Time integration routines for systems of differential equations are essentially a sequence of function compositions involving elementary operations on an input (state) vector, and AD exploits this fact to compute exact derivatives via repeated application of the chain rule \citep{blondel2022efficient,jax2018github}. 
AD is therefore a powerful tool in optimisation problems, and differentiable numerical solvers can be straightforwardly combined with neural network architectures. 
The \texttt{JAX} library \cite{jax2018github} in particular has gained significant popularity, leading to exciting advances in molecular dynamics \cite{jaxmd2020} and numerical methods for computational fluid dynamics \cite{kochkov2021, LCspectral}. 
Gradient based optimisation has already shown great promise as a tool for generating robust guesses for unstable periodic orbits \citep{page2022recurrent}.

In this paper we develop a new AD-based solver for point vortex evolution in a range of geometries, which we combine with a traditional Newton-GMRES-Hookstep algorithm to comprehensively tackle the problems outlined in the recent reviews \cite{Newton2009, Newton2014}.
\REV{Similar to \citet{campbell1979vortex}, we will also primarily focus here on the unbounded domain to both (i) enable a direct comparison to early results and (ii) to provide an efficient way to document solutions without pre-selecting a specific rotation rate.}
We are able to assemble an extremely large (tens of thousands) collection of vortex crystals near the free-energy minima for a range of number of vortices $7 \le N \le 50$, and discover new continuous families of equilibria which were previously unknown. 
Moreover, the flexibility of AD makes it straightforward to compute both dynamical (constant free-energy orbits) and non-dynamical pathways between nearby states to allow for detailed exploration of the free-energy landscape. 

The rest of this work is structured as follows.
In \S\ref{sec:comp_setup}, we formalise the two dimensional point vortex model in a rotating disk, and describe our approach to search for low energy vortex crystals.  
In \S\ref{sec:req}, we assemble large numbers of vortex crystals for numbers of vortices $7 \le N \le 50$ and discuss the new continuous families of double-ringed configurations which emerge. 
In \S\ref{sec:pathways}, we outline and apply an approach to compute energy-minimising pathways between locally stable configurations.  
In \S\ref{sec:connections}, we compute homoclinic orbits for saddles close to the energy minimising states and examine their transient dynamics. 
We conclude in \S\ref{sec:conclusion} with a summary of our results. 

\section{Computational setup \label{sec:comp_setup}}

\subsection{Physical problem}
We consider two-dimensional, irrotational flow in a disk of radius $R$, in which we place $N$ point vortices of equal circulation, $\Gamma$. 
The disk rotates at constant angular velocity, $\Omega$. 
Each point vortex moves at a velocity due to the induced velocity from (1) all of the others and (2) the image vortices.
\REV{The images are required such that the wall-normal induced velocity vanishes at the boundary -- i.e. the (complex) position $|z^*| = R$ is a streamline. So}
if vortex $\alpha$ is located at a dimensional complex position $ z^* = z_{\alpha}^*= x_{\alpha}^*+iy_{\alpha}^*$, then its image sits at $z^* = R^2 / \overline{z}_{\alpha}^*$, where the overbar represents the complex conjugate. 
Lengths are non-dimensionalised with the disk radius, $R$, and time with a reference timescale $R^2 / \Gamma$, 
so that the (dimensionless) evolution equation for each vortex in a frame rotating with the disk is then:
\begin{equation}
    \dot{\overline z}_{\alpha} = -\frac{i}{2\pi}\left(\sum_{\beta=1}^N\phantom{}^{'} \frac{1}{z_{\alpha} - z_{\beta}} - \sum_{\beta=1}^N \frac{1}{z_{\alpha} - 1/\overline{z}_{\beta}}\right) + i \omega \overline{z}_{\alpha},
\label{eqn:full_dynamics}
\end{equation}
where the prime on the summation indicates that $\beta = \alpha$ is excluded and $\omega := R^2 \Omega / \Gamma$ is the non-dimensional disk rotation rate.
Equation (\ref{eqn:full_dynamics}) is equivariant under rotations about the origin of the disk, $\mathscr R^{\theta}: (z_1, \dots, z_N) \to (\exp(i\theta) z_1, \dots \exp(i\theta)z_N)$. 
There is also symmetry under permutation of the vortices, which we must carefully account for when labelling unique configurations or searching for connecting orbits. 

If $\mathbf f^t(\mathbf z)$ is the time forward map of equation (\ref{eqn:full_dynamics}), then equilibria are solutions for which $\mathbf f^t(\mathbf z^*) = \mathbf z^* \; \forall t$. 
These are relative equilibria (REQ) in the lab frame, rotating with the disk at angular velocity $\omega$. 
These REQ, or `vortex crystals', are stationary points of the free energy,
\begin{align}
    \mathcal F := -\sum_{\alpha < \beta} \log |z_{\alpha} - z_{\beta}|^2 &+ \frac{1}{2}\sum_{\alpha, \beta} \log |1- z_{\alpha}\overline{z}_{\beta}|^2 \nonumber \\ &- 2\pi \omega\sum_{\alpha} (1- |z_{\alpha}|^2),
    \label{eqn:free_energy}
\end{align}
which plays the role of a Hamiltonian in the rotating frame.
\REV{Note the absence of a temperature term in the free-energy (\ref{eqn:free_energy}) This would appear in the usual definition of this quantity, but is not required as the results are valid at absolute zero.}
We have scaled the free energy (\ref{eqn:free_energy}) to match the definition in \citet{campbell1979vortex} (note that their `rotation rate' variable is related to ours via $\omega_{CZ} \equiv 2\pi \omega$). 
The equations of motion follow from $\dot{x}_{\alpha} = (1/4\pi) \partial_{y_{\alpha}}\mathcal F $, $\dot{y}_{\alpha} = -(1/4\pi)\partial_{x_{\alpha}}\mathcal F$.
The experimentally observed state in a superfluid is expected to be the global minimiser of $\mathcal F$ \citep{hess1967,campbell1979vortex}. 

At this point, searching for free-energy minimising solutions for a given $N$ also requires the selection of a disk angular velocity, $\omega$, and hence documenting the solutions for a range of $\omega$ becomes a daunting task, \REV{as new solutions continue to emerge as $\omega$ increases}. 
Campbell and Ziff \cite{campbell1979vortex} have shown that the problem is substantially simplified if the image vortices are neglected, which can be justified for `moderate' rotation rates $\omega$ (e.g. at some point beyond the critical $\omega_c$ at which the equilibrium of interest can be maintained in the disk). 
The reasoning is that in order to match an increasing disk angular velocity, the vortices must increasingly bunch together near the origin so that the streamlines of the induced velocity are approximately circular at the boundaries and the images play only a small perturbative role enforcing the no-penetration boundary condition.
In the absence of images, an equilibrium at one value of $\omega$ can then be trivially rescaled to another value via the simple relation $\omega |z_i|^2 = \text{constant}$.
\REV{However, we will see that the number of low-energy crystals grows enormously with increasing $N$, and that the free-energy values become extremely close. 
When considering results it is important to bear in mind that the introduction of the boundaries here may have a significant impact on the ordering of solutions, even if the qualitative effect on the appearance of the crystal is small.
The impact can only be determined in practice by repeating the calculations for a given $\omega$ in system with images (\ref{eqn:free_energy}).}

Without images, the free energy is simply,
\begin{equation}
    \mathcal F_0 = -\sum_{\alpha < \beta} \log |z_{\alpha} - z_{\beta}|^2 - 2 \pi \omega \sum_{\alpha} (1- |z_{\alpha}|^2),
    \label{eqn:F0}
\end{equation}
where the first term is the free-space Hamiltonian, $\mathcal H_0 := -\sum_{\alpha < \beta} \log |z_{\alpha} - z_{\beta}|^2$.
While $\omega$ can now be set arbitrarily, it is helpful to use an $\omega$-independent label to identify and order the relative equilibria. 
A suitable observable is proposed in \citet{campbell1979vortex}, who subtract the free energy of a continuum model, $\mathcal F_c$, evaluated at the same $\omega$.
\REV{The continuum model is derived by developing an approximation to the double sum in equation (\ref{eqn:F0}). To do this, we first note the following integral approximation,
\begin{align*}
    S &:= \sum_{\alpha} \sum_{\beta \neq \alpha}\log 2\pi \omega |\mathbf x_{\alpha} - \mathbf x_{\beta}|^2 \\
    &\approx \frac{N}{\pi r_c^2}\sum_{\alpha} \iint_{|\mathbf x | < r_c} \log 2 \pi \omega |\mathbf x_{\alpha} - \mathbf x|^2 d^2 \mathbf x \\
    &= 2N^2\left(\log \sqrt{2\pi \omega} \,r_c - \frac{1}{4}\right) + N,
\end{align*}
where we have assumed that the vortices are uniformly distributed over a circle of (dimensionless) radius $r_c := r_c^*/R$, and the requirement that the average vorticity of the configuration is twice the rotation rate of the outer cylinder means $r_c = \sqrt{N/2\pi\omega}$.
We then add to this integral approximation a correction term which is the difference between the original sum and the integral, but evaluated for an \emph{infinite} homogeneous vortex lattice,
\begin{equation}
    \Delta S = \sum_{\alpha} \left( \sum_{\mathbf x \in \mathscr L} \log 2\pi \omega |\mathbf x|^2 - 2\omega \iint \log 2\pi \omega|\mathbf x|^2 d^2\mathbf x\right),
    \label{eqn:delta_s}
\end{equation}
where $\mathscr L$ is the set of lattice points under consideration. 
The outer sum in (\ref{eqn:delta_s}) -- which remains over finite $N$ -- is then trivially performed and we write $\Delta S = 2Nb$, where \citet{campbell1979vortex} show that the constant $b\approx 0.74875$ for a triangular lattice.
With that, we write the continuum approximation to (\ref{eqn:F0}) as 
\begin{widetext}
\begin{align*}
    \mathcal F_c(N,\omega) &:= -\frac{1}{2}\left(S + \Delta S\right) + \frac{N(N-1)}{2}\log 2\pi \omega -  2 \pi \omega \sum_{\alpha} (1- |z_{\alpha}|^2) \\
    &= -N\left(b + \frac{1}{2}\right) - \frac{N^2}{2}\left(\log N - \frac{1}{2}\right) + \frac{N(N-1)}{2}\log 2\pi \omega -  2 \pi \omega \sum_{\alpha} (1- |z_{\alpha}|^2),
\end{align*}
\end{widetext}
and finally obtain the $\omega$-independent label for equilibria as
\begin{alignat}{3}
    \Delta f &:= \mathcal F_0(N,\omega)  - \mathcal F_c(N,\omega) & \nonumber \\
    &= -\sum_{\alpha < \beta}\log 2\pi \omega|z_{\alpha} - z_{\beta}|^2 & -\frac{N^2}{4} +\frac{N^2}{2}\log N \nonumber \\ &&+ N\left(b + \frac{1}{2}\right),
    \label{eqn:delta_f}
\end{alignat}
}
which is rotation-rate independent as $\omega|z_i|^2 = \text{constant}$ at equilibrium. 
We focus on the image-less system for the remainder of this paper, apart from \S \ref{sec:valley_of_req} where boundary effects are re-introduced to examine their impact on a certain class of free-space vortex crystals.

\subsection{Relative equilibria guess generation \label{sec:releq_random_search}}
With the removal of images, it is convenient to work in the stationary lab frame, where the vortex velocities can now be simply expressed in terms of the free-space Hamiltonian $\mathcal H_0$:
\begin{equation}
    \dot{x}_{\alpha} = \frac{1}{4\pi}\frac{\partial \mathcal H_0}{\partial y_{\alpha}}, \quad \dot{y}_{\alpha} = -\frac{1}{4\pi}\frac{\partial \mathcal H_0}{\partial x_{\alpha}}.
    \label{eqn:free_space_evoln}
\end{equation}
We solve equation (\ref{eqn:free_space_evoln}) using our fully differentiable point vortex solver \texttt{jax-pv}. 
Time integration is performed with the symplectic second order Runge-Kutta scheme at a fixed time step of $\delta t = 10^{-3}$. 
The numerical solver is built on the \texttt{JAX} library \cite{jax2018github}, which allows for efficient computation of the gradient of the time-forward map $\mathbf{f}^t(\mathbf{x})$, where $\mathbf x := (x_1, y_1, \dots, x_N, y_N)$, with respect to the initial positions of the vortices.
The \texttt{JAX} framework allows us to also leverage efficiency benefits such as just-in-time compilation and auto-vectorisation.  
The ability to minimise loss functions which implicitly depend on the time-forward map is central to this work, and we use it to construct a wide range of REQ candidates which are then converged with a Newton-GMRES-Hookstep solver. 

Our fully differentiable formulation allows for explicit `targeting' of REQ with specific physical properties, and we initially seek candidates by minimising a balance of two loss functions. 
The first contribution acts to minimise the free energy,
\begin{equation}
    \mathcal{L}_{\mathcal{F}} \coloneqq \mathcal{F}_0,
    \label{eq:loss_f}
\end{equation}
while the second demands that the inter-vortex distances $\ell_{\alpha\beta}(t) = \| \mathbf{x}_{\alpha}(t) -  \mathbf{x}_{\beta}(t)\|$ are unchanged after a suitably long simulation:
\begin{equation}
    \mathcal{L}^T_{I} \coloneqq \frac{\mathop{\sum}_{\alpha < \beta} | \ell_{\alpha\beta}(T) - \ell_{\alpha\beta}(0) |^2}{\mathop{\sum}_{\alpha < \beta} | \ell_{\alpha\beta}(0) |^2}.
    \label{eq:loss_releq}
\end{equation}
We combine the loss functions in the following manner:
\begin{equation}
    \mathcal{L} \coloneqq \kappa \mathcal{L}_{\mathcal{F}} + (1 - \kappa)\left(\mathcal{L}^T_{I}+\mathcal{L}^{T/2}_{I}\right),
    \label{eq:loss_random}
\end{equation}
where the contribution from the inter-vortex distances is evaluated at two points along the orbit to encourage fixed inter-vortex distances throughout the computation (constraints at additional times were also tried with minimal impact on performance) and the integration time is fixed at $T=10$.

Prior to optimisation each guess is initialised from a uniform distribution $x_{\alpha}, y_{\alpha} \sim U[0,\sqrt{N}]$. 
We re-centre the configuration such that the centre of vorticity coincides with the origin, and we rescale the configuration throughout optimisation such that its average rotation rate is fixed at $\langle \omega'(t)\rangle = \pi / 2T$. 
The continual updating of $\omega'$ ensures that the pattern traces out a quarter rotation regardless of how the inter-vortex distances are adjusted in gradient descent.
Converged crystals have $\omega |z_i|^2 = \text{constant}$ so can be rescaled arbitrarily and compared to known results using the the $\omega$-independent observable $\Delta f$ (\ref{eqn:delta_f}).

During optimisation the gradient $\boldsymbol \nabla_{\mathbf x(0)} \mathcal L$ is passed to an Adam optimiser \cite{Kingma2015, deepmind2020jax} with initial learning rate $10^{-2}$. 
We then take guesses for which $\mathcal{L}_{I}^{T} \le 10^{-3}$ within a maximum $N_{\text{opt}}=2500$ optimiser steps and attempt to converge them with a Newton-GMRES-Hookstep solver for relative equilibria \citep{Viswanath2007}. 
Over the course of the optimisation procedure we anneal $\kappa$ from 1 to 0 in increments of $1/N_{\text{opt}}$. 
In general, the `good' guess criteria $\mathcal{L}_{I}^{T} \le 10^{-3}$ is achieved while $\kappa$ is still close to 1 (typically after approximately 150--300 optimiser steps),where the loss function (\ref{eq:loss_random}) is dominated by the free energy. 
However, the gradual and slow introduction of the $\mathcal{L}_I$ terms help to reduce the number of optimiser steps required, especially at higher values of $N$, and yields a much richer family of REQ than the $\mathcal L_{\mathcal{F}}$ term alone, which tends to produce the global minimiser.
This can be thought of as an exploration-exploitation trade-off, as in the Reinforcement Learning literature \cite{wang2018}.

At the Newton-GMRES stage, we consider guesses to be converged when 
\begin{equation}
    \frac{\|\mathscr R^{\theta} \mathbf f^T (\mathbf x) - \mathbf x \|}{\|\mathbf x\|} < 10^{-10},
\end{equation}
where $\theta = -\omega T$.  
After the Newton convergence of a large set of vortex crystals, the final step in the process is to extract all unique REQ. 
This must be done carefully to avoid classifying symmetric copies (e.g. under rotation or permutation) as unique crystals. 
When comparing two configurations, both are first scaled to the same value of $\mathcal{H}_0$. 
This is done by scaling one of the configurations $\mathbf x \to \gamma \mathbf x$, with
\begin{equation}
    \gamma = \exp{ \left(\frac{\Delta \mathcal{H}_0 }{ N(N - 1)}\right) },
    \label{eq:same_H_scaling}
\end{equation}
where $\Delta \mathcal{H}_0$ is the difference in Hamiltonian of both configurations. 
Then, all $N(N-1)/2$ inter-vortex distances $\{ \ell_{ij}^k \}_{i=1,\dots,N,j=i+1,\dots,N}$ are computed and sorted for each of the two configurations, where $N$ is the number of vortices in each configuration and $k = 1,2$ is a label for the first and second configuration, respectively. The configurations are classed as symmetric copies if
\begin{equation}
    \frac{| \ell^1_{ij} - \ell^2_{ij} |}{| \ell^1_{ij}|} < 10^{-5}, \quad \forall i,j 
    \label{eq:uniqueness_condition}
\end{equation}
This uniqueness condition is different from the `ring labelling' condition from \citet{campbell1979vortex}, as we do not restrict ourselves to REQ based on concentric rings. 

\subsection{Targeted search}
\label{sec:targetted_search}
As $N$ increases we anticipate the number of REQ will grow dramatically, hence we augment the random search method described above with another approach to search for `missing' low energy states starting from a library of converged solutions.
The appropriate loss function here is
\begin{equation}
    \mathcal{L}_S \coloneqq \mathcal{L}_I^T + \nu \sigma\left(\frac{\mathcal{F} - \mathcal{F}^*}{\delta} \right),
    \label{eq:loss_sigmoid}
\end{equation}
which seeks to converge onto REQ lower than some target free energy $\mathcal{F} < \mathcal{F}^*$ via the sigmoidal contribution $\sigma(\bullet)$, which (approximately) will `fire' if $\mathcal F > \mathcal F^*$. 
We set the hyperparameter $\nu=100$, the large value favouring lower energy states.
We apply a recursive approach to optimise for (\ref{eq:loss_sigmoid}), beginning from a converged solution, $\mathbf x^*$, at $\mathcal F = \mathcal F^*$, where we select $\mathcal F^*$ to coincide with the free energy of a previously converged state.
Typically we do this for all states with a corresponding $\Delta f$ below some threshold value $\Delta f_{\text{max}}$. 
We take the converged solution and perturb each vortex position according to $(x_{\alpha}, y_{\alpha}) \to (x_{\alpha} + \varepsilon x_{\alpha}', y_{\alpha} +\varepsilon y_{\alpha}')$, where $\varepsilon = 0.1$ and $x_{\alpha}', y_{\alpha}' \sim U[0,1]$. 
We do this 20 times to generate 20 new candidate equilibria, before then performing gradient descent on (\ref{eq:loss_sigmoid}) with an initial $\delta = 10^{-1}$. 
For this REQ data set augmentation we find that the simpler AdaGrad optimiser \citep{Duchi2011} is particularly effective. Our initial learning rate here is $\eta = 0.1$. 
Since we begin this optimisation process with a small perturbation to a converged equilibrium, the initial value of the inter-vortex loss (our usual convergence metric) is typically $\mathcal{L}_{I}^{T} \lesssim 10^{-5}$. 
To ensure that we generate new candidates and not simply repeats of our starting solution, we require that the optimiser performs at least 25 steps. 
During this process the loss component $\mathcal{L}_{I}^{T}$ increases while the vortex positions adjust to reduce the large sigmoidal contribution to $\mathcal L_S$.
We pass solutions for which $\mathcal{L}_{I}^{T} \le 10^{-4}$ to a Newton solver. 
Finally, new unique solutions with $\Delta f < \Delta f_{\text{max}}$ are identified via (\ref{eq:uniqueness_condition}). 

\section{Vortex crystals in the image-less system \label{sec:req}}
\subsection{Crystal distributions}
\begin{figure}
  \centering
  \includegraphics[width=\linewidth]{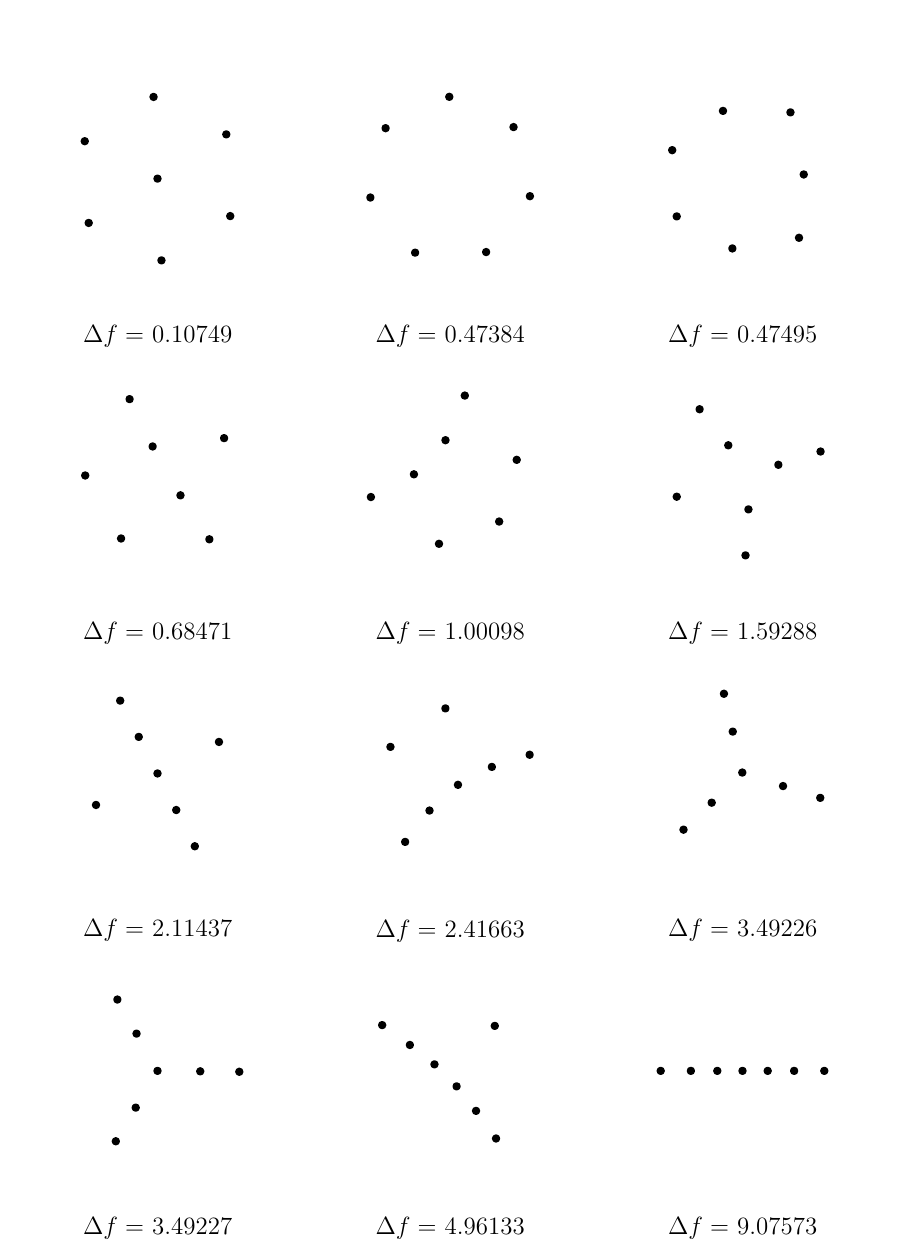}%
  \caption{All relative equilibria for the point vortex system with $N = 7$, in order of increasing $\Delta f$. 
  \label{fig:n7_all_states}}
\end{figure}
We begin with an initial test case at $N=7$, where it has been rigorously established that there are exactly 12 REQ \citep{faugere2012}. 
In this case we seek to converge all known solutions and hence do not wish to restrict ourselves to low free energies, so we use a version of loss function (\ref{eq:loss_random}) with $\kappa = 0$ throughout optimisation. 
The `random search' approach yields 11 of the 12 known solutions, with only the high-$\mathcal F$ co-linear state requiring special attention -- we find this solution by initialising our vortices on a straight line -- presumably because its basin of attraction in the optimisation is not favoured by the random vortex initialisations.
The REQ solutions and corresponding free-energy labels are reported in figure \ref{fig:n7_all_states}. 
To generate the first 11 solutions we started with 1000 random initialisations, from which we converged 188 states in total. 
The most frequently computed REQ was the energy minimising crystal at $\Delta f = 0.10749$ (54 times), while the REQ at $\Delta f = 4.96133$ was only computed once. 
Setting $\kappa = 0$ is an important modification when attempting to compute REQs at high energies:
if $\kappa$ is annealed as previously described, then we are unable to find the crystal at $\Delta f = 3.49227$, even after 8000 random initialisations. 
This is likely due to the very similar structure at the slightly lower value of $\Delta f = 3.49226$, which is preferred by the $\mathcal L_{\mathcal F}$ term in (\ref{eq:loss_random}).

\begin{table*}   
  \caption{Summary of computations for various numbers of vortices. Note that the success rate shown in brackets in the columns is defined relative to the raw number in the previous column. The success rate in the final column should be interpreted differently to those preceding it -- a low success rate here for a large number of samples suggests that most, or potentially all, unique states may have been obtained.} 
  \label{tab:rel_eq_randomsearch}
  \begin{ruledtabular}
  \begin{tabular}{c|c|c|c|c}
    N & \# samples & \# AD candidates (success rate) & \# Newton convergences (success rate) & \# unique REQ (success rate) \\ \hline
    10 & 9000 & 8845 (0.983) & 8796 (0.994) & 28 (0.003) \\
    11 & 2550 & 2504 (0.982) & 2245 (0.897) & 30 (0.013) \\
    12 & 50000 & 41425 (0.828) & 27611 (0.667) & 49 (0.002) \\
    13 & 2550 & 2500 (0.98) & 1517 (0.607) & 65 (0.043) \\
    14 & 2550 & 2510 (0.984) & 1343 (0.535) & 83 (0.062) \\
    15 & 2550 & 2503 (0.982) & 1996 (0.797) & 138 (0.069) \\
    16 & 31400 & 30840 (0.982) & 19733 (0.64) & 265 (0.013) \\
    17 & 2550 & 2501 (0.981) & 2059 (0.823) & 256 (0.124) \\
    18 & 5050 & 4954 (0.981) & 3326 (0.671) & 409 (0.123) \\
    19 & 15000 & 14708 (0.981) & 10427 (0.709) & 851 (0.082) \\
    20 & 37200 & 36472 (0.98) & 14237 (0.39) & 1728 (0.121) \\
    30 & 102600 & 100474 (0.979) & 44258 (0.44) & 38577 (0.872) \\
    50 & 65700 & 64052 (0.975) & 16150 (0.252) & 15521 (0.961) \\
  \end{tabular}
  \end{ruledtabular}
\end{table*}

We now consider large numbers of guesses for $N = 10, 11, \dots, 20$, as well as $N=30$ and $N=50$. 
A summary of our initial random search using loss function (\ref{eq:loss_random}) is reported in table \ref{tab:rel_eq_randomsearch}. 
We searched extensively at $N \in \{10, 12, 16, 20, 30, 50\}$, while other vortex numbers were considered to explore the possible appearance of degenerate families of vortex crystals that we discuss later in \S\ref{sec:valley_of_req}. 
Notably, we are able to generate very large numbers of viable guesses via automatic differentiation (success rates at this stage are almost all $\gtrsim 0.98$).
As is often found in turbulence \citep{Gibson2008,Chandler2013}, the success rate of the Newton algorithm is a bottleneck, particularly at higher $N$, although even at $N=50$ we are still able to converge around a quarter of our AD guesses. 


\begin{figure*}
  \centering
  \includegraphics[width=\linewidth]{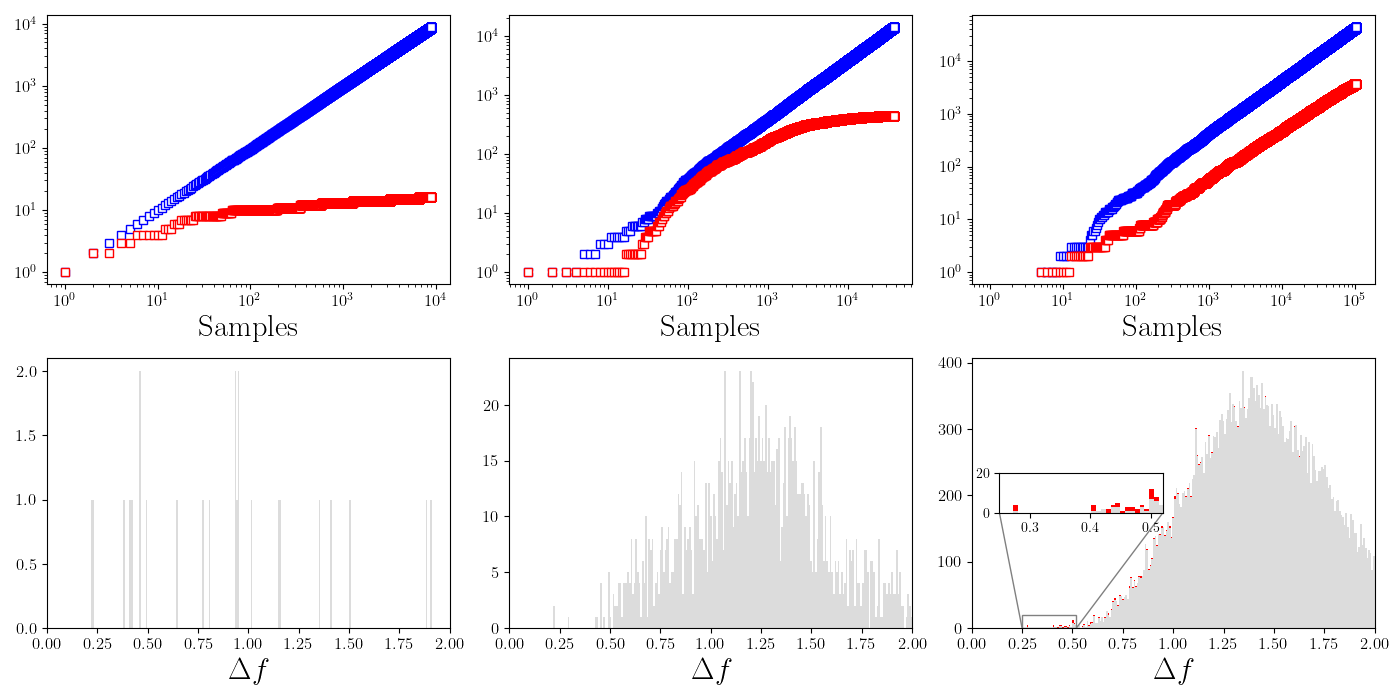}%
  \caption{(Top) Unique convergences below $\Delta f = 1$ (red) and the number of raw convergences including repeats (blue) as a function of the number of samples. (Bottom) Histograms of the distribution of states as a function of the free energy with bin width $\Delta f = 0.08$. Results are shown for numbers of vortices $N=10$ (left), $N=20$ (centre) and $N=30$ (right). The red contributions to the distribution at $N=30$ indicates extra solutions found by applying the targeted search for low energy REQs using equation (\ref{eq:loss_sigmoid}).} 
  \label{fig:rel_eq_efficiency}
\end{figure*}
The number of unique convergences identified (as classified by equation (\ref{eq:uniqueness_condition})) is also described in table \ref{tab:rel_eq_randomsearch}. 
These numbers should be interpreted in the context of the raw number of samples considered for each $N$.
For example, the large numbers of guesses considered at $N=10$ or $N=12$ and the very low `success rate' in the unique solutions column indicates that we may have found most (potentially all) REQ in those cases, while the high success rates at $N=30$ and $N=50$ indicates that we have found only a small subset of all available crystals. 

We examine the converged solutions further in figure \ref{fig:rel_eq_efficiency}, where we plot both the number of Newton convergences and the number of unique convergences below $\Delta f = 1$ against the number of samples over the course of the computation for $N \in \{10, 20, 30\}$. 
At both $N=10$ and $N=20$ the number of unique convergences is found to level off while the number of Newton convergences remains roughly constant. 
The plateauing of the number of unique solutions is observed for roughly an order of magnitude more samples when $N=20$ compared to the $N=10$ case. 
At higher $N=30$ we continue to observe roughly constant linear growth in the number of unique solutions even at $O(10^5)$ guesses. 

We also explore the distribution of states in as a function of $\Delta f$ for $N\in\{10, 20, 30\}$ in the lower panels of figure \ref{fig:rel_eq_efficiency}.
Clearly the density of states increases enormously with increasing $N$, and by $N=30$ there is some indication that there is an exponential increase in the number of REQ with increasing free energy.
Note that the explicit search for low free-energy solutions in our loss (\ref{eq:loss_random}) is likely the reason for the sudden drop in numbers of convergences beyond $\Delta f \approx 1$. 

Due to the indications in table \ref{tab:rel_eq_randomsearch} and the top-right panel of figure \ref{fig:rel_eq_efficiency} that we are far from identifying all REQ at $N=30$, we supplement our random search there with the targeted sweep described in section \ref{sec:comp_setup}C. 
We set $\Delta f_{\text{max}} = 0.5$, below which we initially had 19 converged solutions (see inset in figure \ref{fig:rel_eq_efficiency}).
Each of these 19 solutions is used to define a value of $\mathcal F^*$ in loss function (\ref{eq:loss_sigmoid}), and we generate 20 perturbed states for each, or 680 new guesses in total. 
Of these 680 guesses, 190 unique vortex crystals were converged, of which 17 were new solutions for which $\Delta f < \Delta f_{\text{max}}$.
Additional solutions found this way are indicated in red on the histogram in figure \ref{fig:rel_eq_efficiency}; the low-energy sweep was necessary to find the lowest free-energy state, which differs from that reported previously \citep{campbell1979vortex}.
Note that the Newton solver also produces a large number of states for which the converged $\Delta f > \Delta f_{\text{max}}$ -- see the red detail in figure \ref{fig:rel_eq_efficiency}.

\begin{figure*}
  \centering
  \includegraphics[width=\linewidth]{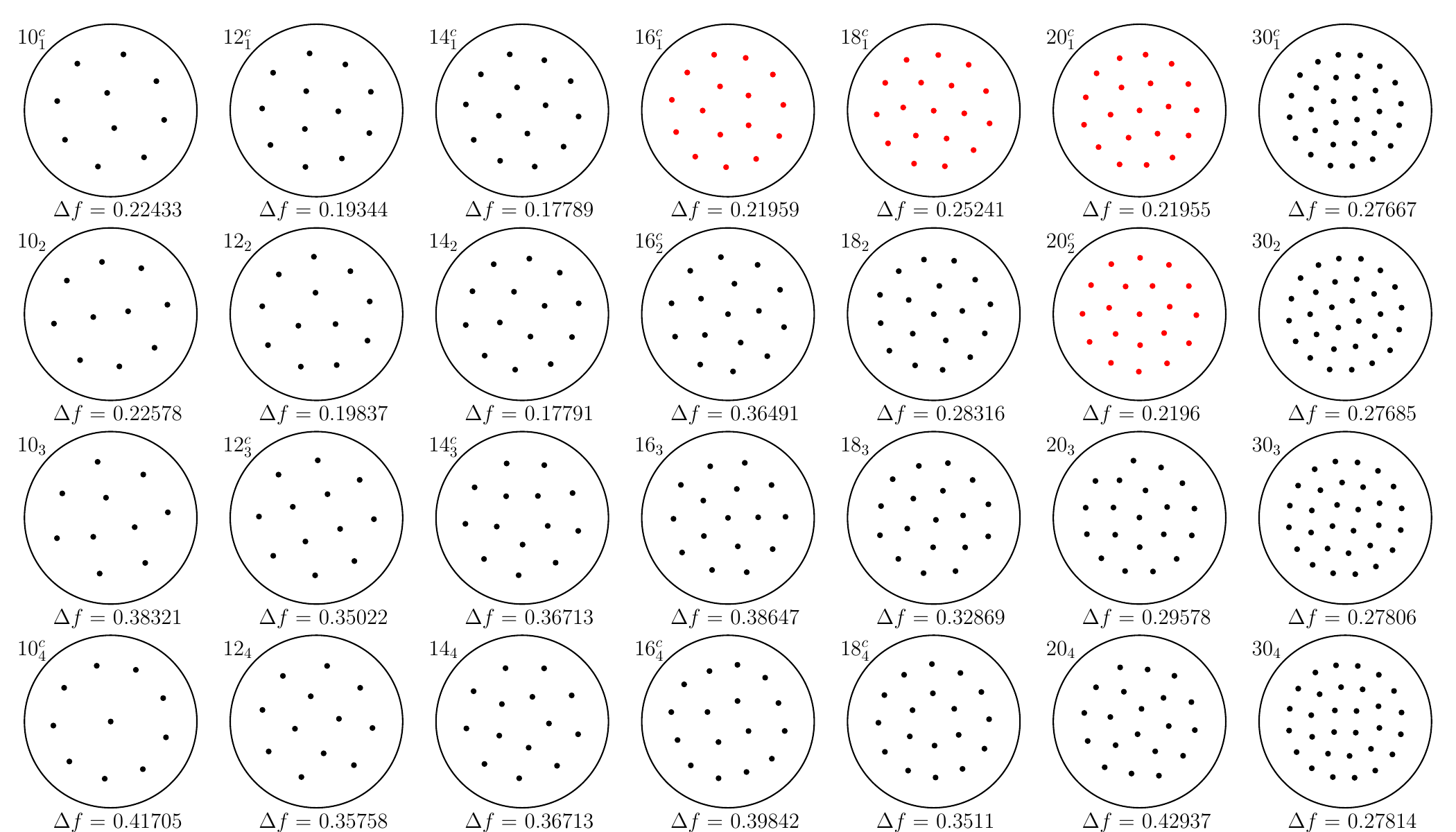}%
  \caption{The four energy minimising states for $N = \{10, 12, 14, 16, 18, 20, 30 \}$ returned from the random search method detailed in \S \ref{sec:releq_random_search} (with additional states found at $N=30$ via the targeted search described in \S \ref{sec:targetted_search}). 
  The solutions plotted in red denote members of continuous solution families (see \S\ref{sec:valley_of_req}). 
  The states are labelled by their number of vortices, with a subscript denoting their ordering by $\Delta f$, while superscript `c' is used to identify centres. 
  Solutions $10_3$, $14_2$, $18_3$, $20_3$, $20_4$ and \emph{all} crystals at $N=30$ shown here are new (not listed in \citet{campbell1979vortex}), including a new global minimiser $30_1^c$.}
  \label{fig:min4_n10to_30}
\end{figure*}
We report the four energy minimising states for $N \in \{10, 12, 14, 16, 18, 20, 30 \}$ in figure \ref{fig:min4_n10to_30}.
This figure includes some states identified by \citet{campbell1979vortex}, but with many additional solutions. 
In figure \ref{fig:min4_n10to_30} there are new states at $N \in \{10,14,18,20,30\}$ (identified in detail in the figure caption). 
Most notably, \emph{all} four solutions at $N=30$ in the figure are new, including a new global minimum of the free energy.

A handful of solutions shown in figure \ref{fig:min4_n10to_30} are stable centres (including, as expected, the energy minimising states), though the majority are saddles. 
The stability of all states is determined by computing the eigendecomposition of the Jacobian of the vortex velocities (the right hand side of equation (\ref{eqn:full_dynamics}) without images), in a frame rotating at angular velocity $\omega$ with the crystal. 
Most of the saddle solutions in the vicinity of the free-energy minimum have low dimensional unstable manifolds.
Stability properties of the 10 lowest free-energy solutions for all $N$ considered are reported in appendix \ref{app:detail}, along with additional details of the states themselves.
All states possess a neutral eigenvector (eigenvalue $\lambda_0=0$) corresponding to a rotation through the origin. 

For some values of $N$ we observe a large number of `unique' states (as identified by the symmetry-invariant observable (\ref{eq:uniqueness_condition})) with identical values of $\Delta f$. 
These distinct states all possess an additional neutral eigenvector, and appear to be smoothly connected to one another -- i.e. given any two crystals with equal $\Delta f$ which are `close', we can always interpolate to find another intermediate equilibrium between them. 
Vortex crystals exhibiting this behaviour are shown in red in figure \ref{fig:min4_n10to_30}, where we are actually reporting only one of a large number of converged solutions at that particular $\Delta f$. 
To our knowledge these continuous families of vortex crystals have not been documented before, and we now explore them in more detail. 

\subsection{Degenerate double ring solutions}
\label{sec:valley_of_req}
\begin{figure}
  \centering
  \includegraphics[width=\linewidth]{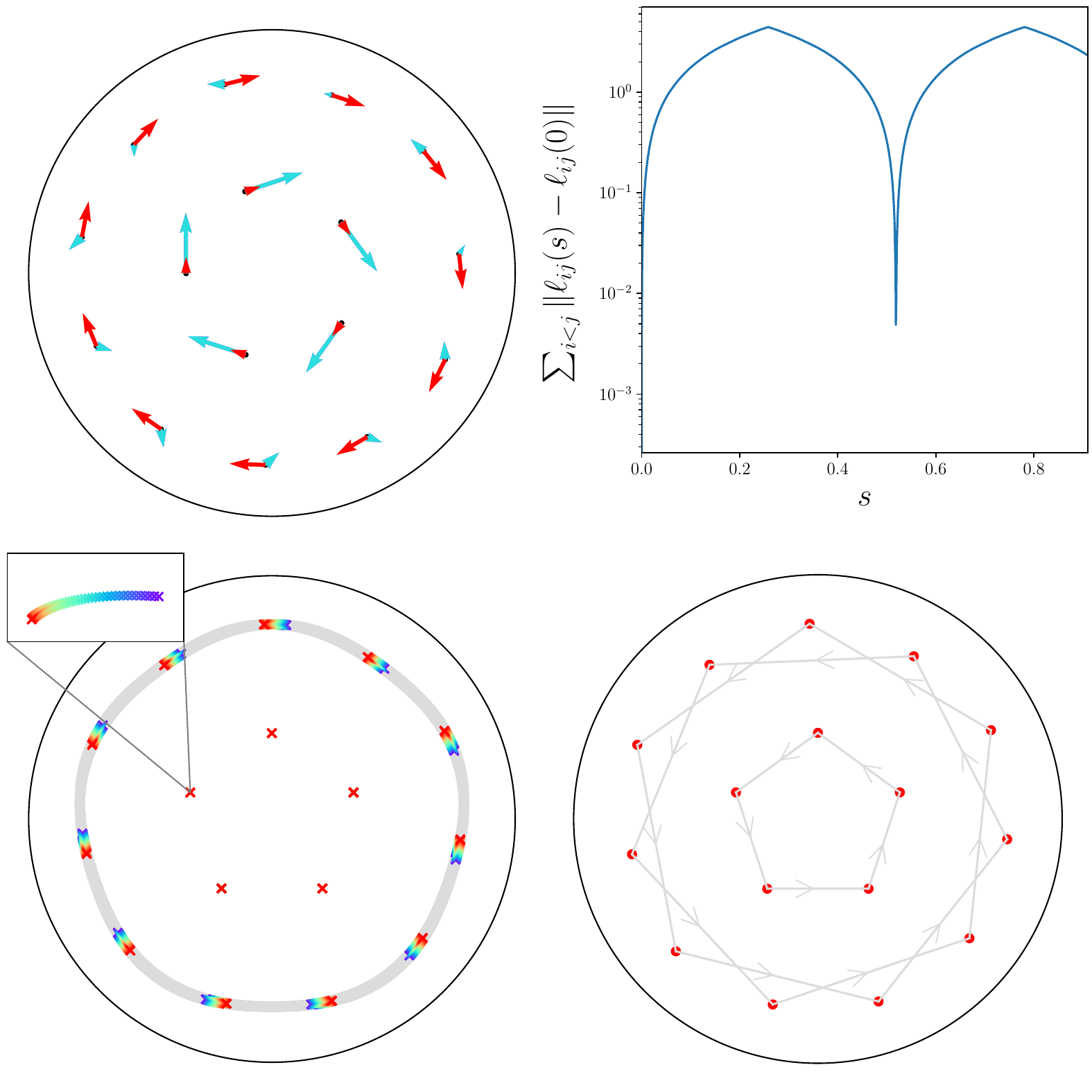}%
  \caption{Continuous family of vortex crystals at $N=16$ (the free-energy minimiser). (Top left) The two neutral eigenvectors overlayed on one state from the solution family. Red arrows correspond to the continuous rotational symmetry while the blue arrows correspond to the new non-trivial direction. (Top right) Sum of the difference of intervortex distances between a crystal in the family and a starting state as a function of arclength, $s$. The valley has a periodicity of approximately $\Delta s \approx 0.5$ and we start from a mirror-symmetric state (see red markers in bottom panels). (Bottom left) The movement of the vortices along one cycle ($\Delta s \approx 0.5$) through the family is shown in colour. For ease of visualisation we fix the location of one of the inner ring vortices as we move through the family. Note the small motion of the inner ring highlighted in the inset. The grey outline in the background shows a continuation of the family over many cycles, and traces out a pentagon shape to match the vortices on the inner ring. (Bottom right) Arrows show the permutation of vortices that has occurred after one cycle.\label{fig:n16_family_plot}}
\end{figure}
We examine both `neutral' eigenvectors ($\lambda_0^{(1,2)} = 0$) for the energy minimising state at $N=16$ in figure \ref{fig:n16_family_plot}, which is the lowest $N$ at which we found a continuous family of solutions.
In the top left panel of that figure we see that one of these eigenvectors corresponds to the rotational symmetry in the system (red arrows) as expected.
The second eigenvector, which is identified with blue arrows, is markedly different:
in the co-rotating frame it appears that the inner ring rotates while the outer ring wobbles in a non-axisymmetric manner.
The dependence of the free energy on the logarithm of the relative distances between vortices does in principle allow for more complex behaviour than uniform rotation/translation:
Consider a continuous family of solutions parameterised by some arc-length $s$ (formally, a group orbit). 
The condition that $\partial_s \Delta f =0$ is equivalent to
\begin{equation}
    \sum_{i<j} \frac{1}{\ell_{ij}(s)}\frac{d \ell_{ij}(s)}{ds} = 0.
    \label{eq:crystal_constraint}
\end{equation}
Clearly, this condition holds for pure rotation as $d_s \ell_{ij} =0 \; \forall i, j$, but non-trivial combinations of vortex displacements are also allowed. 

As a further confirmation of a continuous family of crystals, we also verified that the second directional derivative of the free energy along the relevant neutral eigendirection was zero. 
In the free-energy landscape, we can think of the family of solutions as a smooth, continuous valley along which the crystal structure varies in a non-trivial manner (in contrast to the rotational symmetry). 
Computationally, we are able to take any member of a continuous family of vortex crystals and move smoothly through the valley by perturbing it along the relevant neutral eigendirection, $\mathbf x^* \to \mathbf x^* + \varepsilon \hat{\mathbf x}_0'$. 
We set $\varepsilon = 5\times 10^{-4}$ and construct a sequence of connected states $\{ \mathcal P \mathbf{x}^*_i\}$ by re-converging the perturbed solution with a Newton solver before repeating the process, performing a pullback \cite{Fels1998, SIMINOS2011187, budanur2015} $\mathcal P$ along the rotational symmetry direction (see appendix \ref{app:pullback}) to account for any drift in the rotational-symmetry valley introduced by the Newton solver.
The connected sequence of states is parameterised by an arclength, $s_n := s_{n-1} + \|\mathcal P \mathbf x_{n}^* - \mathcal P \mathbf x_{n-1}^*\|_2$, and we stop the calculation once we reach a symmetrically equivalent copy of the starting solution.

The family of solutions obtained via arclength continuation at $N=16$ is reported in the remaining panels of figure \ref{fig:n16_family_plot}. 
In the top-right panel of the figure we measure the sum of the differences in inter-vortex distances, relative to the initial state, identifying a `period' of $\Delta s \approx 0.5$.
Note that we start this continuation from a state with mirror symmetry (see red markers in lower panels of figure \ref{fig:n16_family_plot}). 
The vortex configurations that make up one arclength-period of the continuous family are visualised in the lower panels of figure \ref{fig:n16_family_plot}, where we choose a reference frame in which the angle of one of the inner-ring vortices is fixed. 
We observe that the outer ring vortices move smoothly relative to the inner ring until a permuted copy of the starting solution is reached. 
The permutations are shown in the final panel of the figure: the vortex labels of the inner ring are rotated through one `click' while the outer ring vortices are permuted by two `clicks'. 
Note that the vortices on the inner ring are slightly perturbed as we move through the family. 
If the arclength process is continued through many cycles, the outer ring of vortices is observed to move through a smoothed approximation to a pentagon -- matching the number of vortices on the inner ring -- which is shown with the thick grey line in the bottom left panel of figure \ref{fig:n16_family_plot}. 

\begin{figure}
  \centering
  \includegraphics[width=\linewidth]{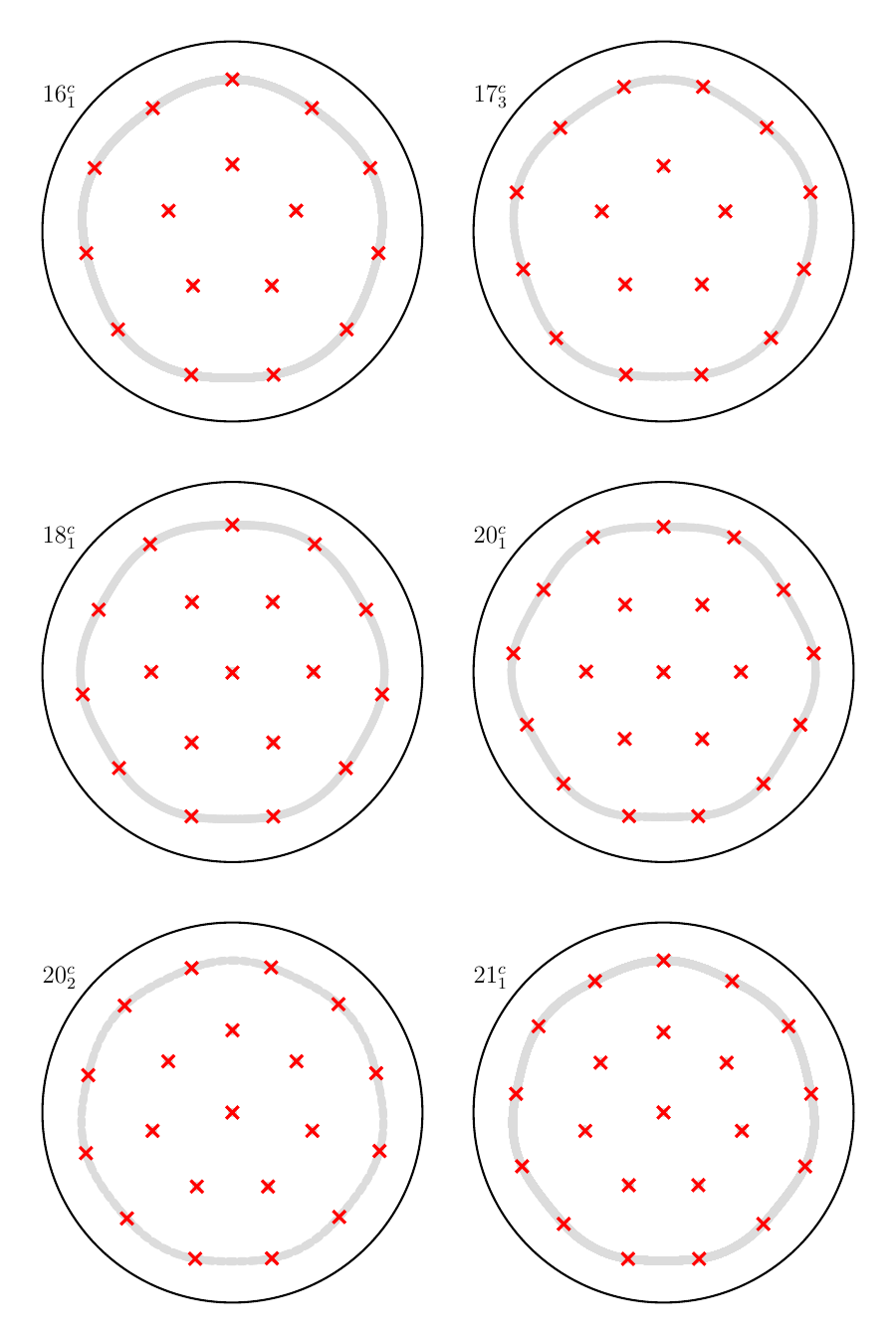}%
  \caption{Continuous energy minimising valleys for $N \in \{16,17,18,20 ,21\}$. These were the only states exhibiting this behaviour in our solution library. Here we show in red a mirror symmetric state in the family, along with the states reached via arclength continuation in grey. Note the tendency of the outer rings to approximately follow a geometric shape set by the number of vortices on the inner ring. In all examples we use a reference frame in which the location of one of the inner ring vortices is fixed.  \label{fig:all_degenerate_families}}
\end{figure}
We searched our library of solutions for other examples of continuous families of crystals, and all configurations exhibiting this behaviour are summarised in figure \ref{fig:all_degenerate_families}.
The states are always `double ring' configurations, and there may or may not be a central vortex. 
In all cases, following the continuous valley of solutions results in the outer ring moving relative to the inner on a smooth approximation to a geometric shape specified by the number of vortices on the inner ring. For example, if there are six vortices on the inner ring, the outer vortices approximately move around a hexagon. 

\begin{figure}
  \centering
  \includegraphics[width=\linewidth]{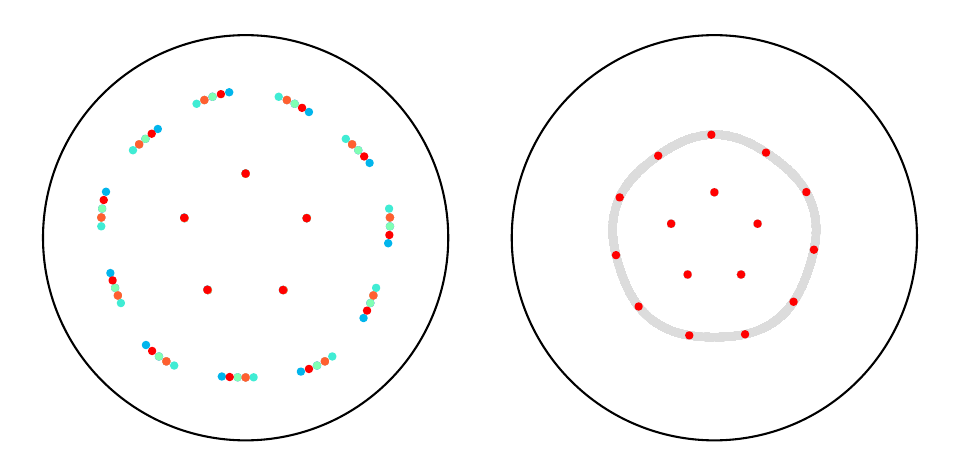}%
  \caption{Examples of the discrete, unique energy minimising equilibria at $\omega_{CZ} = 20$ (left) and $\omega_{CZ} = 40$ (right). The contraction of the vortices towards the axis of rotation for the higher rate of rotation is evident when comparing the two plots. All 9 unique equilibria at $\omega_{CZ} = 20$ are shown in various colours. These 9 equilibria are arranged in 5 `bands'; any equilibria in the same band are very close to overlapping. All 1128 unique equilibria computed at $\omega_{CZ} = 40$ are shown in grey, illustrating the approach to the unbounded limit as $\omega_{CZ}$ is increased.}
  \label{fig:discrete_valley}
\end{figure}
The continuous families of solutions identified for certain double ring configurations exist for the unbounded problem (no images).
It is natural to query what happens when the images are included \REV{due to the sudden requirement that $|z|=1$ is a streamline}.
\REV{This is anticipated to be of significant importance} at lower rotation rates, where the proximity of the image vortices will have a non-negligible impact in the problem.
To examine the importance of the images we apply the random search algorithm from \S \ref{sec:releq_random_search} to target strict equilibria in the non-inertial frame. 
As the vortices are now constrained to a unit radius disk, the guesses are initialised from a uniform distribution $x_{\alpha}, y_{\alpha} \sim U[-0.7,0.7]$. 
As $\omega$ is increased, the vortices in a free-energy minimising equilibrium are brought closer to the axis of rotation so that the induced velocities are sufficiently high to match the imposed rotation rate.
To compensate for these larger induced velocities, and to be consistent with our scaling in the unbounded domain, we set the integration time $T = 2 \pi/\omega$. 
We also fixed the annealing parameter $\kappa = 1$ as we were only interested in computing the energy minimiser in the rotating disk domain. 
Otherwise, the random search algorithm and optimisation parameters are unchanged from \S \ref{sec:releq_random_search}. 

In the full problem with images, we find that the continuous families are replaced by a discrete set of states with equal free energy.
For example at $N = 16$ with $\omega_{CZ}=20$  (the critical rotation rate is $19 < \omega_{CZ}^c < 20$\citep{campbell1979vortex}) we converged 64 energy minimising solutions from 1000 samples, of which 9 were unique solutions. 
Increasing slightly to $\omega_{CZ}=22$, we converged 36 energy minimising equilibria from the same number of samples, of which 12 were unique solutions. 
The very low rate of \emph{unique} convergences suggests that we may have found all unique equilibria at these rotation rates.
At a much higher value $\omega = 40$, the images have a much smaller impact in the problem and we find a very high rate of unique convergences, generating 2020 energy minimising equilibria from 4250 samples, of which 1128 are unique solutions.
The states found at $\omega_{CZ}=20$ and $\omega_{CZ}=40$ are summarised in figure \ref{fig:discrete_valley}.
At the lower rotation rate the outer ring vortices for the various configurations appear to be approximately uniformly spaced and arranged in a series of five `bands' while some locations (relative to the inner ring) are completely inaccessible.
At the higher rotation rate we obtain a very close approximation to the continuous family reported in figure \ref{fig:n16_family_plot}.


\section{Energy-minimising pathways between relative equilibria}
\label{sec:pathways}
In a number of cases we have found that the free-energy minimising solution could actually be any one of a continuous family of states -- e.g. this is true for all $N$ considered in figure \ref{fig:all_degenerate_families} apart from the 17 vortex crystal. 
Experimental realisations of these configurations \citep{yarmchuk1979,BELattice2001,Grzybowski2000} typically show a sequence of unstable states en route to the minimising solution, or even transient behaviour at long time \citep{Grzybowski2000}. 
It is natural to question whether there is a `preferred' state within the continuous family which is more likely to be observed in an experiment. 
More generally, it would be useful to map out and visualise the free-energy landscape around local minima, as global minima with a particularly shallow free-energy barrier may not be robust to small perturbations to the system, while a local minimum with high barrier energy might be observed in practice. 
To explore the free-energy landscape around a particular vortex crystal, we adapt a form of the doubly nudged elastic band (DNEB) method \cite{trygubenko2004}, to find non-dynamical, free-energy minimising pathways between REQ.  
This method has been shown to work successfully in multiple settings, such as structural self assembly \cite{goodrich2021} and rearrangements of the Lennard-Jones cluster \cite{trygubenko2004}. 
The DNEB methodology should be contrasted with the simple approach adopted by \citet{campbell1979vortex}, who searched for minimising pathways with the constraint that the pathway from one state to the other coincides with a monotonic outward radial motion of one vortex from an inner ring to an outer ring, which we will see is often not the case.

To approximate the energy-minimising pathway between two REQs, $\mathbf{x}_a^*$ and $\mathbf{x}_b^*$, the DNEB method seeks an energy-minimising chain of $N_I$ intermediate states $\{ \mathbf{x}_i\}_{i=1,\dots,N_I}$. 
For the point vortex system in the unbounded domain, we define a `potential energy' as the sum of the free energies along the chain:  
\begin{equation}
    V := \sum_{i=1}^{N_I} \mathcal{F}(\mathbf{x}_i).
\end{equation}
Minimising $V$ alone would likely result in states `sliding down' to one of the two minima \cite{trygubenko2004}, and to counter this effect we add a fictional elastic band potential which inserts frictional high-dimensional springs between adjacent states: 
\begin{equation}
    \Tilde{V} := \frac{1}{2} k \sum_{i=1}^{N_I+1} \mathcal L_G(\mathbf x_i, \mathbf x_{i-1}),
\end{equation}
where $\mathbf{x}_a^* \equiv \mathbf{x}_0$ and $\mathbf{x}_b^* \equiv \mathbf{x}_{N_I+1}$. 
For standard Hookean springs, we would write $\mathcal L_G (\mathbf x_i, \mathbf x_{i-1}) = \| \mathbf x_i - \mathbf x_{i-1}\|^2$; in the vortex pathway problem it is necessary to modify this term to account for both the rotational and permutation symmetries in the system, which we describe below. 
Energy minimising pathways are then determined by minimising an overall loss function
\begin{equation}
    \mathcal L_P := V + \Tilde{V}. 
    \label{eqn:path_loss}
\end{equation}

To account for the permutation symmetry we define a permutation-invariant observable by centering a two-dimensional Gaussian field over each vortex. 
For an $N$-vortex configuration $\mathbf x^*$, this observable is defined as
\begin{equation}
    G(x, y; \mathbf x^*) = \frac{1}{2\pi \sigma^2}\sum_{\alpha=1}^N \exp \left(\frac{-(x - x^*_{\alpha})^2 - (y - y^*_{\alpha})^2}{2\sigma^2}\right).
    \label{eq:gaussian_smear}
\end{equation}
The hyperparameter $\sigma$ is set as $\sigma = 10^{-2} A /N$, where $A=L^2$ is the area of the computational grid on which $G$ is evaluated; typically we set $L = 1.25\max{|\mathbf x^*|}$ and evaluate $G$ on a square grid with $N_x=N_y=64$ grid points. 
\begin{figure}
  \centering
  \includegraphics[width=\linewidth]{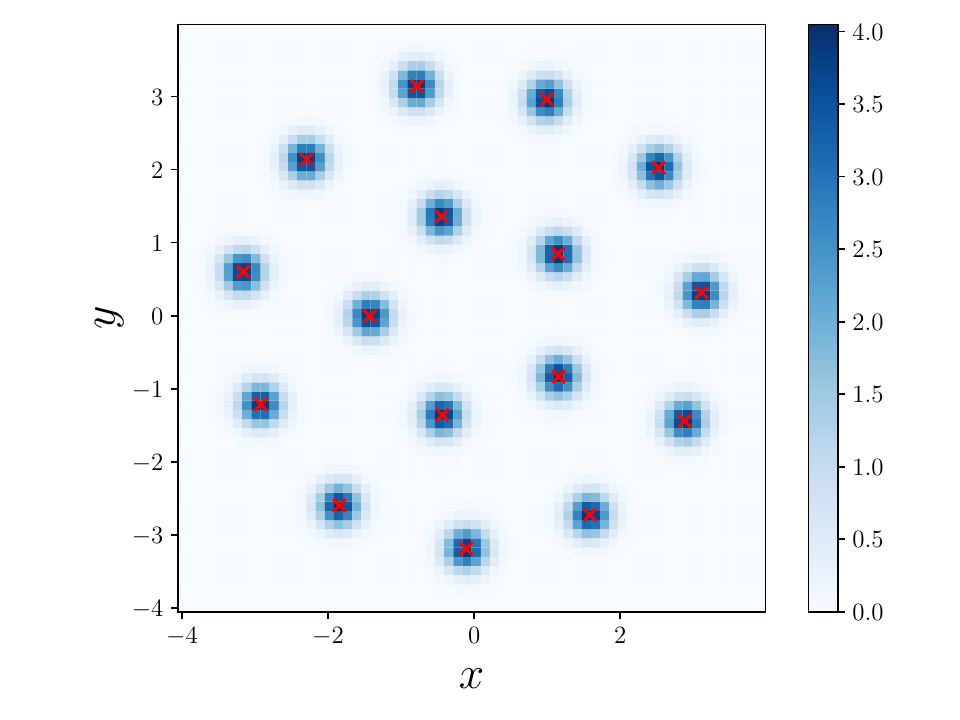}%
  \caption{Permutation-invariant Gaussian smearing of the energy minimiser at $N=16$, scaled to rotate at $\omega = \pi/2T$ with $T=10$. }
  \label{fig:n16_gaussiansmear_eg}
\end{figure}
An example of observable (\ref{eq:gaussian_smear}) is reported in figure \ref{fig:n16_gaussiansmear_eg} for the $N=16$ system.
The vortex configuration is scaled so that it rotates at $\omega = \pi/2T$ with $T=10$ to match our convention from \S \ref{sec:releq_random_search}. 
The Gaussian observable is then used to define the spring term appearing in equation (\ref{eqn:path_loss}):
\begin{align}
    \mathcal L_G &(\theta, \mathbf x_i, \mathbf x_{i-1}) := \nonumber 
    \\
    &\frac{1}{L_x L_y} \iint \left(G(x,y; \mathbf x_i) - G(x, y; \mathscr R^{\theta} \mathbf x_{i-1})\right)^2 dx dy,
\end{align}
where the state $\mathbf x_{i-1}$ has been rotated through some angle $\theta$.
The angle used in between each pair of states on the chain is determined simply via 
\begin{equation}
    \theta^*_i = \argmin_{\theta} \mathcal L_G(\theta; \mathbf x_i, \mathbf x_{i-1}).
    \label{eqn:inner_opt_chain}
\end{equation}

Putting this all together the overall loss function for finding minimum energy pathways between states (\ref{eqn:path_loss}) reads 
\begin{align}
    \mathcal L_P(\{\mathbf x_i\}_{i=1}^{N_I}, &\{\theta^*_i\}_{i=1}^{N_I}) = \nonumber \\
    &\sum_{i=1}^{N_I} \mathcal{F}(\mathbf{x}_i) + 
    \frac{1}{2}k \sum_{i=1}^{N_I+1} \mathcal L_G(\theta_i^*, \mathbf x_i, \mathbf x_{i-1}).
    \label{eqn:vortex_final_chain_loss}
\end{align}
The two states $\mathbf{x}_a$ and $\mathbf{x}_b$ on either side of the chain are kept fixed throughout the optimisation process. 
We solve the outer optimisation problem (\ref{eqn:vortex_final_chain_loss}) using an AdaGrad optimiser with a fairly aggressive initial learning rate $\eta = 0.5$, and classify our chain as converged when the relative difference between sequential iterations of the optimiser was less than $10^{-6}$. 
At each outer optimisation step we must also solve the inner optimisation problem (\ref{eqn:inner_opt_chain}) for the crystal angles, which is done in $O(10)$ steps with an Adam optimiser (learning rate $\eta = 10^{-2}$. 
Throughout we set the spring constant to $k=10^{-2}$ --larger values caused the intermediate states close to the REQs to clump together-- and the number of interpolating states to $N_I = 200$.
Further details of the optimisation, particularly how gradients are used in the DNEB method, are provided in appendix \ref{app:dneb}. 
To initialise the chain we linearly interpolate between the two vortex crystals $\mathbf x_a^*$ and $\mathbf x_b^*$, where we rotate $\mathbf x_b^*$ such that it maximally overlaps with $\mathbf x_a^*$. This is done as a pre-processing step rather than as part of the chain optimisation, similar to the approach in \citet{goodrich2021}. 

\begin{figure*}
  \centering
  \includegraphics[width=\linewidth]{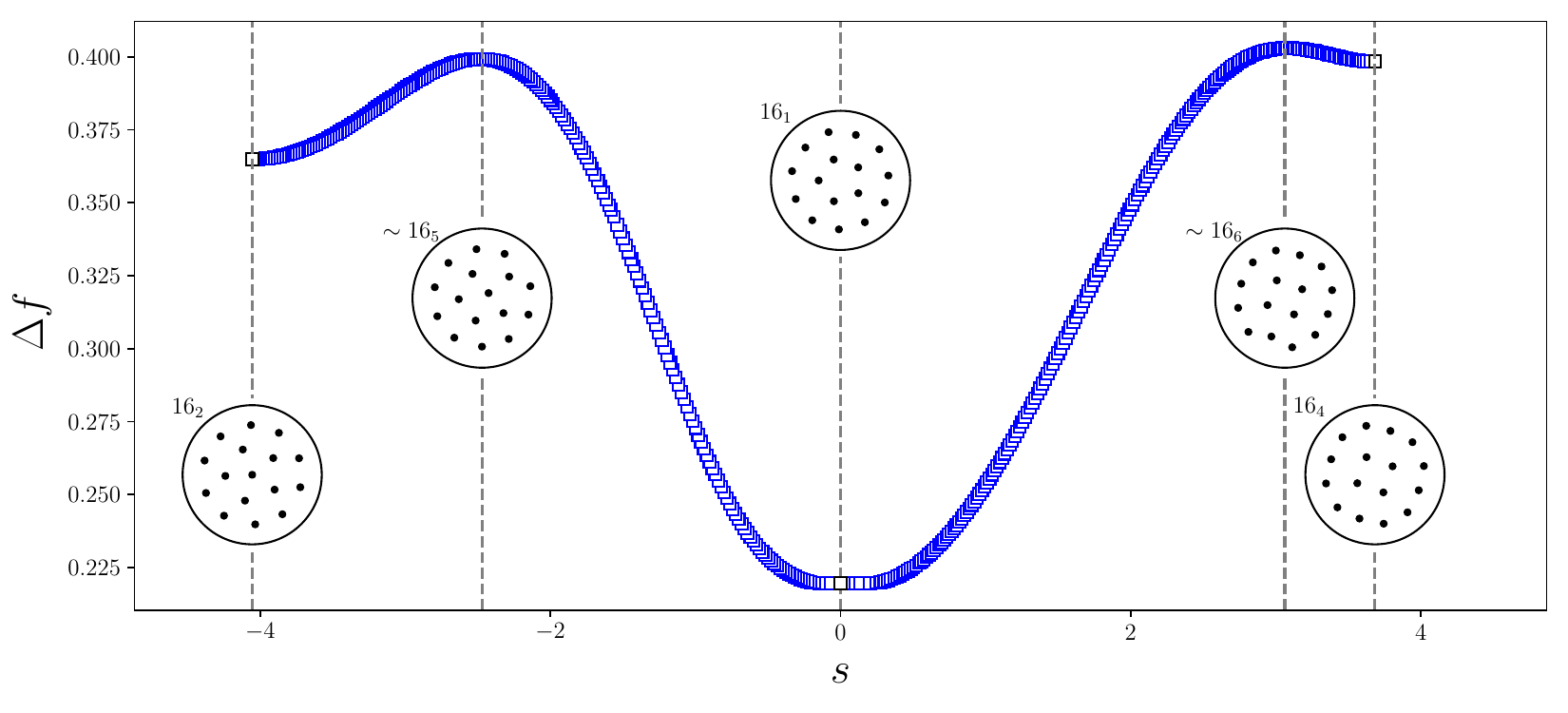}%
  \caption{
  Energy minimising pathways from local minima in the $N=16$ system into the global minimum. Two pathways are shown here on the left and right of the figure respectively, where the the centres correspond to solutions reported in figure \ref{fig:min4_n10to_30}. Both pathways move through a saddle point which is also a state identified in the random search (see details of these solutions in appendix \ref{app:detail}). }
  \label{fig:pathways_to_family}
\end{figure*}
\begin{figure}
  \includegraphics[width=\linewidth]{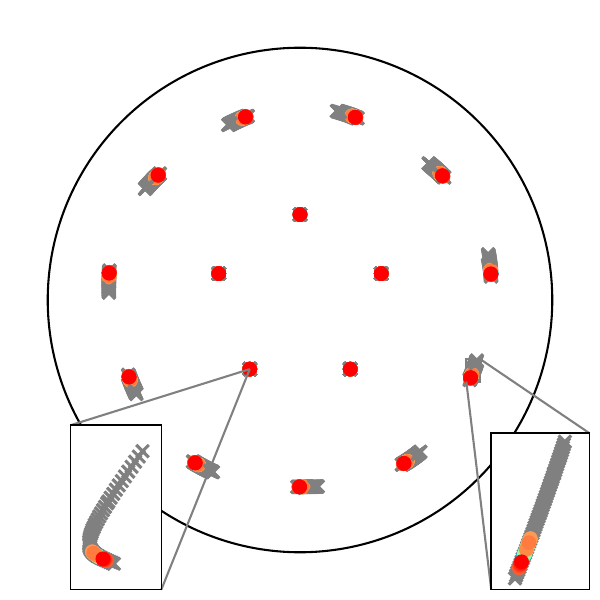}%
  \caption{
  Common entry point (colours) into continuous family of solutions (grey), defined via a tolerance of $10^{-5}$ on the relative difference between $\Delta f$ of the entry point and $\Delta f(16_1^c)$. 
  The grey crosses identify 50 equally-spaced equilibria from with the continuous family of solutions that defines REQ $16_1^c$. An energy-minimising pathway was computed from REQ $16_4$ to each. The non-dynamical pathway always approaches the configuration identified in red (which is a mirror symmetric state) before moving along the valley to the respective final state in grey. The clustering of the entry points for each pathway (coloured from blue to red0 is so tight that only the orange- and red-coloured states are visible. }
  \label{fig:n16_pathways_preferred_entry}
\end{figure}
We now apply the methodology outlined above to find minimising pathways into REQ $16_1^c$ in the $N=16$ case, looking for pathways from the two next lowest energy centres (see figure \ref{fig:min4_n10to_30}). 
These pathways are reported in figure \ref{fig:pathways_to_family}. 
This case was also considered in \citet{campbell1979vortex} but their pathways featured discontinuities in the free energy due to the constraint on how the optimisation was performed, i.e. radial motion of a vortex between rings.
No such discontinuities are found here, and the smooth routes from the nearby local minima $16_2^c$ and $16_4^c$ both pass through unstable REQ (these states on the chain were converged in a couple of Newton steps). 
The barrier energy from $16_2^c$ is considerably higher than that from $16_4^c$ (note the correspondence of the pattern $16_2^c$ with that seen in self assembly experiments of \citet{Grzybowski2000}). 

The state at $16_1^c$ is the simplest example of a degenerate continuous family of solutions discussed in \S\ref{sec:valley_of_req}.
While one example from the family is considered in figure \ref{fig:pathways_to_family}, we can use our methodology to determine whether there is a preferred state when computing energy minimising chains from nearby local minima. 
To do this, we initialise 50 states uniformly along one period of the valley of solutions (see discussion in \S\ref{sec:valley_of_req}) and compute pathways from the local minimum $16_4$ to these 50 states. 
The 50 `final' states from the solution valley are shown in grey in figure \ref{fig:n16_pathways_preferred_entry}. 
We then assess whether the pathway has entered the solution valley before continuing along the constant free-energy surface to the final state. 
We determine an `entry point' by selecting the last interpolation state on the pathway which is within a relative difference of $10^{-5}$ of $\Delta f(16_1^c)$, using the Newton solver to then converge this state to a REQ. 
The 50 entry points determined this way are shown in colour for all pathways considered in figure \ref{fig:n16_pathways_preferred_entry}, the tight clustering of points clearly indicating a preferred state.
If no preferred entry state existed, we would expect the coloured entry states to be spread evenly along the grey final states. Instead, the coloured entry states are clustered around one preferred entry state. The entry states become even more tightly clustered around the preferred entry state if the tolerance on the relative difference of $\Delta f(16_1^c)$ is increased to $10^{-4}$. 
\begin{figure}
  \centering
  \includegraphics[width=\linewidth]{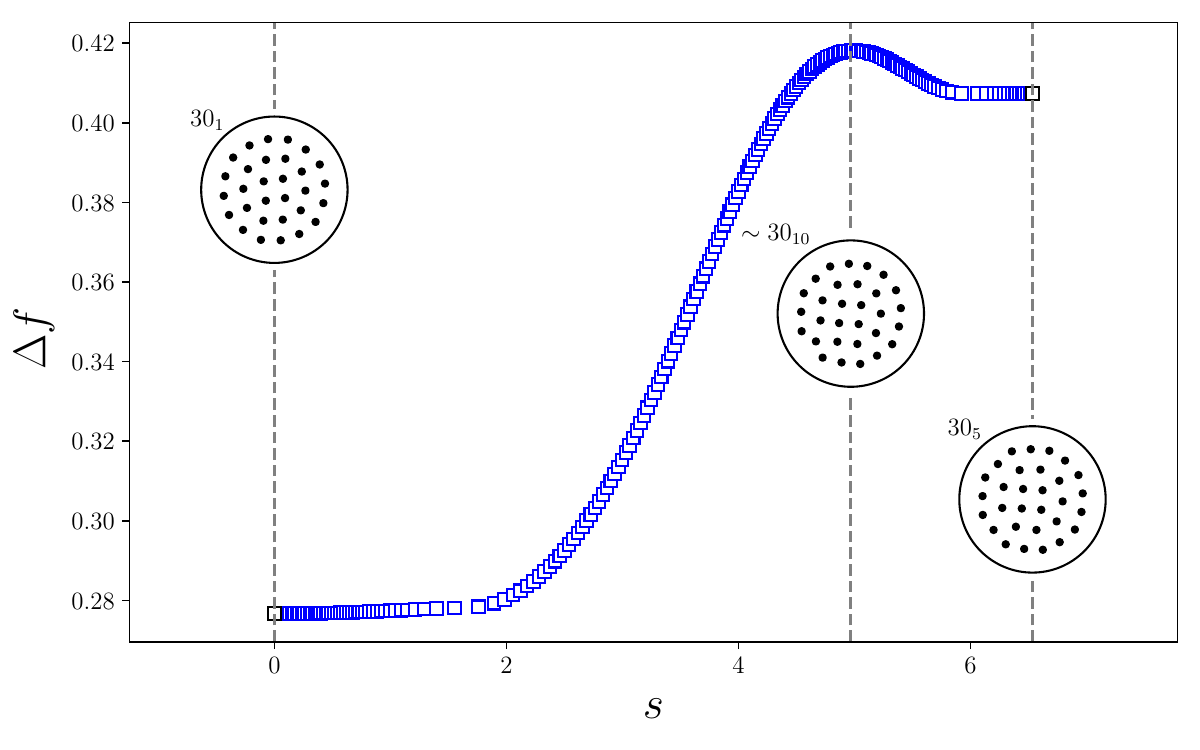}%
  \caption{Energy minimising pathway at $N = 30$ pathway from the next lowest energy centre REQ $30_5^c$ to the minimiser $30_1^c$. A saddle point from our library of solutions is again found along the pathway (see details of these solutions in appendix \ref{app:detail}).}
  \label{fig:n30_pathway}
\end{figure}
The energy pathways optimisation does not involve any dynamics and is easily deployed at much higher $N$. 
For example, we consider a pathway between centres in the $N=30$ system in figure \ref{fig:n30_pathway}, which as expected passes through an unstable saddle before a very slow descent into $30^c_1$.
We can adapt the methodology here to explore the dynamics connected to the weakly unstable saddles that deform the free-energy landscape around the local minimiser; modifying the Gaussian-based loss (\ref{eqn:path_loss}) to find homoclinic orbits, exploiting the ability of the AD implementation to efficiently differentiate through trajectories.

\section{Homoclinic connections}
\label{sec:connections}
Unlike convergence of many simple invariant solutions (equilibria, REQ, periodic orbits) there is no generally adopted methodology for computing heteroclinic or homoclinic orbits.
So-called shooting methods have found some connections in plane Couette flow \cite{halcrow2009}, while a recently-proposed variational method has been used to compute connections for the one-dimensional Kuramoto-Sivashinsky equation \cite{ashtari2023}.
Here, we outline an optimisation method exploiting a fully differentiable solver.

\subsection{Candidate orbit generation}
Generally speaking, we find only a single REQ at any specific energy level (see table \ref{tab:rel_eq_library} in appendix \ref{app:detail}), hence we restrict our search to homoclinic orbits. 
To generate candidate trajectories we perform a simple recurrent flow analysis\citep{Chandler2013} to identify returns to a starting equilibrium $\mathbf x^*$.
Given the large number of vortices we anticipate that connections are likely to be between permuted copies of $\mathbf x^*$, hence we use the permutation invariant observable (equation \ref{eq:gaussian_smear}) and modify the associated loss so that 
\begin{equation}
    R(\mathbf x_0, t) := \min_{\theta} \mathcal L_G (\theta, \mathbf f^t(\mathbf x_0), \mathbf x^*)
\end{equation}
is our measure of similarity used to flag possible connecting orbits.

We seed the recurrent flow analysis calculations with initial conditions $\mathbf x_0 = \mathbf x^* + \varepsilon \mathbf x'$, where $\mathbf x' \in E^u(\mathbf x^*)$ (the unstable subspace of the REQ being considered) and we write
\begin{equation}
   \mathbf{x}'(\bm{c}) =  \sum_{i=1}^{N_U} c_i \bm{v}_i,
\end{equation}
where $\{\bm{v}_i\}$ are unstable eigenvectors of $\mathbf x^*$, before normalising $|\mathbf x'| = 1$.
We set the constant $\varepsilon = 10^{-5}$. 
A coarse initial search with the constants $c_i \in \{0, \pm 1\}$ is performed, and we use the same discretisation grid for $G$ as described in \S\ref{sec:pathways}. 
We searched over a time horizon $t\in [0, 1000]$, computing $R(\mathbf x_0, t)$ every time unit. 
Cases where $R \leq 0.1$ are saved as candidates for connecting orbits.

\subsection{Convergence}
Suitable guesses are converged using gradient-based optimisation. 
First, a modified Jonker-Volgenant variant \cite{crouse2016} of the Hungarian algorithm \cite{kuhn1955} is used to compute vortex permutations at the end of the connecting orbit. 
The Hungarian algorithm is a combinatorial optimisation algorithm that solves the linear assignment problem in $O(N^3)$. 
The linear assignment problem in this setting is
\begin{equation}
    \tilde{\mathbf P} = \argmin_{\mathbf P} \text{trace} ( \mathbf P \mathbf C ),
\end{equation}
where $\mathbf P$ is a permutation matrix which permutes the rows of a matrix $\mathbf C$ whose elements $C_{ij}$ are the Euclidean distances between the $i^{th}$ vortex of the initial REQ state $\mathbf{x}^*$ and the $j^{th}$ vortex in the final state on the connecting orbit candidate $\mathbf {f}^t(\mathbf{x}_0)$. Then, for the suitable candidates, $\mathbf f^t(\mathbf x_0) \approx \tilde{\mathbf P}\mathbf x^*$. 

Once we have identified a candidate connection and determined the required vortex permutations, the second stage is to converge the connecting orbit. 
As the permutation symmetry is now respected, we minimise the Euclidean loss function:
\begin{equation}
    \mathcal{L}^t_{E}\left( \bm{c}, \theta \right) :=  \left\| \mathscr R^{\theta} \mathbf{f}^t(\mathbf{x}_0(\bm{c})) - \tilde{\mathbf P} \mathbf{x}^*  \right\|^2_2.
    \label{eq:euclidean_connection_loss}
\end{equation}
The idea here is that we must remain in the unstable subspace of $\mathbf{x}^*$ by fixing the perturbation size $\varepsilon$, but the specific direction is unknown. 
We minimise (\ref{eq:euclidean_connection_loss}) at fixed integration time $t$ (discussed further below) using an Adam optimiser with an initial learning rate of $\eta_c= 10^{-3}$ to determine the vector of constants $\bm{c}$, and an Adam optimiser with an initial learning rate of $\eta_{\theta}=10^{-2}$ to determine the optimal $\theta$. 

The initial integration time $t$ is set by the criteria $R \le 0.1$ from the candidate orbit generation step.
We then carefully increase $t$ throughout the convergence process. Each time $t$ is increased, the connecting orbit is converged to a smaller tolerance level. 
In practice, we increase the integration time by $1\%$ 10 times, while logarithmically decreasing the tolerance from $10^{-3}$ to $10^{-5}$. 
We allow a maximum number of 1000 optimiser steps every time we converge the connection to a smaller tolerance. 
When $\mathcal{L}^t_{E} < 10^{-5}$, we consider the connection converged, which we confirm by verifying that the dynamics are linear in the vicinity of $\mathbf x^*$.

\begin{figure}
  \centering
  \includegraphics[width=\linewidth]{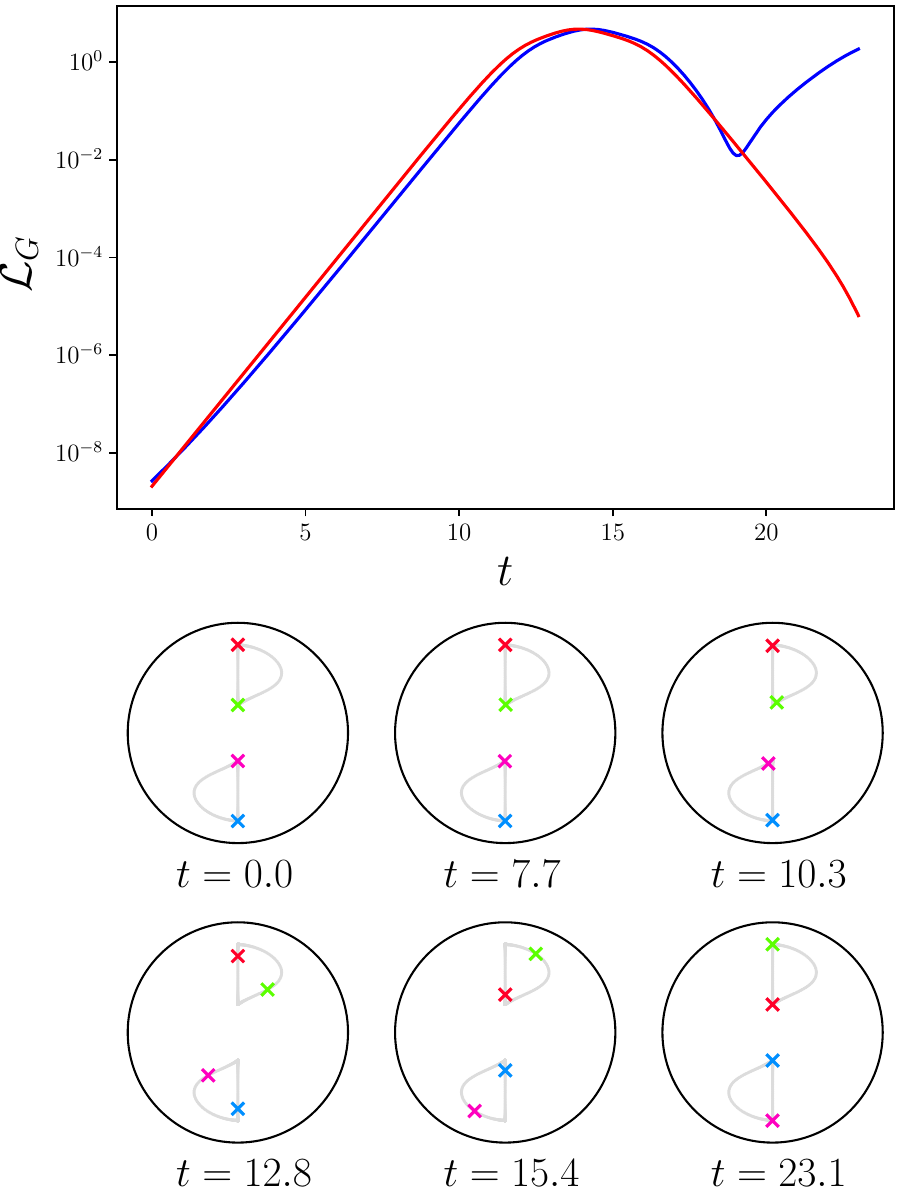}%
  \caption{(Top) The evolution of $\min_{\theta}\mathcal{L_{G}}$ over approximately 23 time units for the 4 vortex system, for both the candidate connection (blue) and the converged homoclinic connection (red). (Bottom) Snapshots at various stages of the converged homoclinic connection in a co-rotating frame. The result of the connection is that the inner vortices are swapped with the outer vortices.} 
  \label{fig:n4_connection}
\end{figure}
As a proof of concept, we successfully computed the only connection in the integrable 3-vortex system \cite{makarov2021}, which is a homoclinic connection between the collinear state via a permutation of vortices. 
We then found a homoclinic connection in the non-integrable 4-vortex system, also between two permuted collinear states (with a two-dimensional unstable subspace).
This connection is reported in figure \ref{fig:n4_connection}, where we show both the initial evolution of $\mathcal L_G$ that flagged the guess along with the `converged' $\mathcal L_G$. The vortex evolution is shown in a co-rotating frame, the two (initially) inner-most vortices moving outwards along straight lines as the two outer vortices loop back to replace them. 
\begin{figure}
  \centering
  \includegraphics[width=\linewidth]{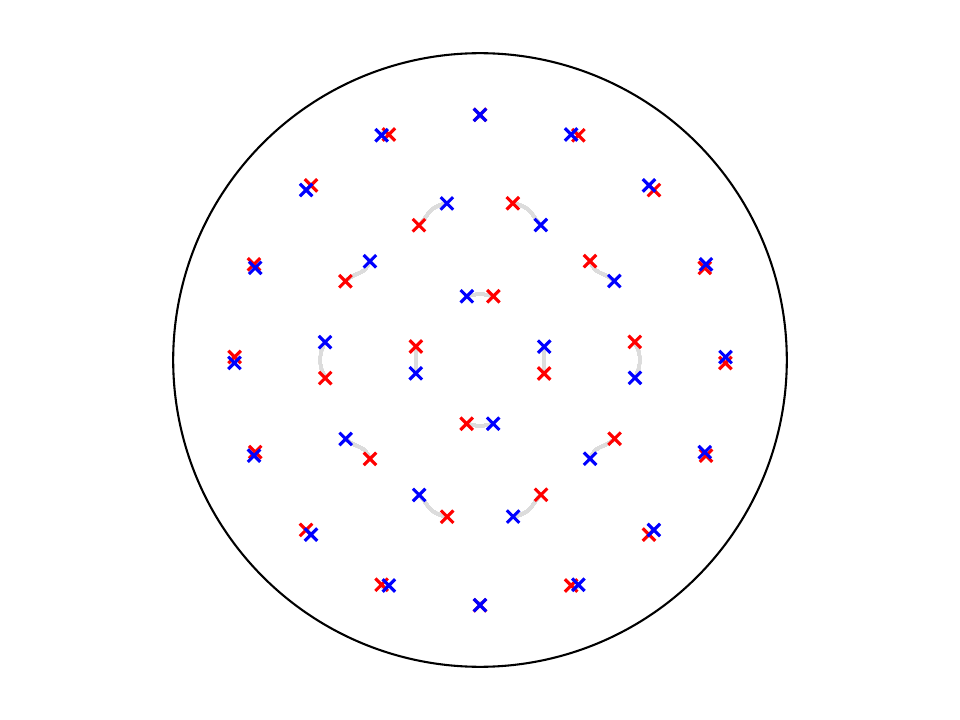}%
  \caption{A homoclinic connection for REQ $30_4$ is achieved via a permutation of vortices, and is shown here in a rotational symmetry reduced frame. The connection evolves (grey outline) from REQ $30_4$ (red crosses) to its permuted copy (blue crosses). }
  \label{fig:n30_connection}
\end{figure}
Many of unstable saddles reported in \S\ref{sec:releq_random_search} (see also appendix \ref{app:detail}) have low dimensional unstable manifolds, and often feature along free-energy minimising pathways between minima as observed in \S\ref{sec:pathways}. 
The method outlined above for homoclinic orbits gives us a further tool with which to explore the range of dynamical events we may expect to see in a dissipative relaxation onto the free-energy minimising state. 
We report one such example in figure \ref{fig:n30_connection}, where we find a homoclinic orbit for a weakly unstable saddle (REQ $30_4$) in the $N=30$ system, with a 2-dimensional unstable subspace.

\section{Conclusion}
\label{sec:conclusion}
In this paper we have introduced a new computational approach to efficiently identify large numbers of relative equilibria in point vortex dynamics, focusing on free-energy minimising states in a rotating cylinder. 
Using a combination of modern optimisation techniques with a fully differentiable solver, we were able to assemble tens of thousands of low-energy vortex crystals for a wide range of vortex numbers $N$, with some evidence that we may have found \emph{all} low-energy states for lower values $N\lesssim 20$. 

As part of the search for energy minimisers we discovered a new set of double-ringed continuous families of vortex crystals in unbounded domains, in which the outer ring can be smoothly deformed relative to the inner. 
These crystals are often the energy minimiser when they occur. 
The introduction of a solid wall breaks up the continuum of solutions into a set of discrete states with identical free energies, though the number of such states increases dramatically as the disk rotation rate is increased and the role of the image vortices diminishes.

Finally, we developed computational methods to explore the free-energy landscape in the vicinity of the global minimiser, finding both non-dynamical, minimum-energy pathways and dynamical homoclinic orbits for unstable saddles. 
When the energy minimiser is a continuous family of states, the non-dynamical pathways from nearby local minima indicate a common entry point into the family, and are suggestive of a preferred state experimentally.
Future work could exploit the differentiability of the solver here to directly connect simple invariant solutions for point vortices to more complex data e.g. to assess the role of vortex crystals in the decay of two-dimensional turbulence \citep{jimenez_2020}.
\REV{Another possibility is to adapt the connecting orbit search to more complex dissipative systems like the Navier-Stokes equations -- while the problem considered in this paper is dissipation-less, none of the trajectory-based methodology relies on this feature.}

\appendix

\section{Vortex crystal details}
\label{app:detail}
In table \ref{tab:rel_eq_library} we report details of the 10 lowest energy states converged for all $N$ considered (not including the $N=7$ test case or the $N=50$ calculations). 
Eigenvalues are normalised by the rotation rate of the configuration. 
In order to provide a useful means of describing and identifying the patterns in table \ref{tab:rel_eq_library}, ring numbers are given for each state. 
We follow the convention in Campbell and Ziff \cite{campbell1979vortex} that two vortices are said to fall in the same ring if their radii agree within 2\%.

\begin{table*}  
  \caption{The 10 lowest energy REQ computed for systems with up to 30 vortices. The results from a sigmoid search are included for the 30 vortex system, otherwise the results come purely from the random search algorithm. \label{tab:rel_eq_library}}
  \begin{ruledtabular}
  \begin{tabular}{cccccc|cccccc}
N & $\Delta f$ & Rings & $N_U$ & $\lambda_{u}$ & $\sum \lambda_{u}$ & N & $\Delta f$ & Rings & $N_U$ & $\lambda_{u}$ & $\sum \lambda_{u}$ \\ \hline 
$10_{1}$ & 0.224329 & (2, 4, 4) & - & - & - & $16_{1}$ & 0.219593 & (5, 11)  & - & - & - \\
$10_{2}$ & 0.22578 & (2, 2, 4, 2)  & 1 & 0.00456 & 0.00456 & $16_{2}$ & 0.364912 & (1, 5, 10)  & - & - & - \\
$10_{3}$ & 0.383211 & (1, 2, 2, 5)  & 1 & 0.00521 & 0.00521 & $16_{3}$ & 0.386468 & (1, 5, 5, 5) & 1 & 0.00696 & 0.00696 \\
$10_{4}$ & 0.417052 & (1, 9) & - & - & - & $16_{4}$ & 0.398422 & (4, 4, 8)  & - & - & - \\
$10_{5}$ & 0.418514 & (1, 1, 2, 2, 2, 2)  & 1 & 0.00798 & 0.00798 & $16_{5}$ & 0.399251 & (1, 2, 2, 1, 10) & 1 & 0.01111 & 0.01111 \\
$10_{6}$ & 0.459785 & (1, 2, 2, 2, 3)  & 2 & 0.01592 & 0.02423 & $16_{6}$ & 0.402959 & (2, 2, 1, 3, 6, 2) & 1 & 0.00937 & 0.00937 \\
$10_{7}$ & 0.461326 & (1, 2, 2, 1, 4)  & 2 & 0.01702 & 0.02828 & $16_{7}$ & 0.405414 & (4, 8, 4) & 1 & 0.00621 & 0.00621 \\
$10_{8}$ & 0.495497 & (1, 2, 1, 2, 4)  & 2 & 0.01530 & 0.02812 & $16_{8}$ & 0.407486 & (1, 1, 2, 1, 2, 2, 7) & 1 & 0.00527 & 0.00527 \\
$10_{9}$ & 0.648659 & (1, 2, 2, 1, 2, 2)  & 2 & 0.01962 & 0.03688 & $16_{9}$ & 0.409302 & (1, 2, 1, 2, 1, 2, 7) & 1 & 0.00741 & 0.00741 \\
$10_{10}$ & 0.77648 & (4, 6)  & 3 & 0.02101 & 0.04161 & $16_{10}$ & 0.412404 & (1, 1, 2, 2, 1, 2, 7) & 1 & 0.00585 & 0.00585 \\
$11_{1}$ & 0.24918 & (3, 3, 2, 3) & - & - & - & $17_{1}$ & 0.283943 & (1, 5, 11)  & - & - & - \\
$11_{2}$ & 0.249185 & (3, 2, 2, 4)  & 1 & 0.00048 & 0.00048 & $17_{2}$ & 0.283943 & (1, 5, 11) & 1 & 0.00003 & 0.00003 \\
$11_{3}$ & 0.288585 & (2, 3, 2, 4) & - & - & - & $17_{3}$ & 0.294121 & (5, 12)  & - & - & - \\
$11_{4}$ & 0.288585 & (2, 2, 2, 2, 3)  & 1 & 0.00012 & 0.00012 & $17_{4}$ & 0.348941 & (2, 1, 2, 1, 7, 4)  & - & - & - \\
$11_{5}$ & 0.323309 & (2, 1, 2, 2, 2, 2)  & 1 & 0.01219 & 0.01219 & $17_{5}$ & 0.349039 & (1, 2, 1, 2, 8, 3) & 1 & 0.00140 & 0.00140 \\
$11_{6}$ & 0.567528 & (1, 2, 1, 7)  & 2 & 0.01087 & 0.01326 & $17_{6}$ & 0.355255 & (6, 11) & 1 & 0.00732 & 0.00732 \\
$11_{7}$ & 0.577026 & (1, 1, 2, 2, 1, 2, 2)  & 2 & 0.01869 & 0.03243 & $17_{7}$ & 0.355255 & (6, 11) & 2 & 0.00732 & 0.00733 \\
$11_{8}$ & 0.58481 & (1, 2, 4, 4)  & 2 & 0.01317 & 0.02477 & $17_{8}$ & 0.357567 & (1, 2, 1, 1, 1, 1, 10) & 1 & 0.00837 & 0.00837 \\
$11_{9}$ & 0.590779 & (2, 2, 1, 6)  & 2 & 0.01838 & 0.03037 & $17_{9}$ & 0.358299 & (1, 2, 1, 2, 1, 8, 2) & 2 & 0.00858 & 0.01160 \\
$11_{10}$ & 0.624168 & (1, 3, 2, 5)  & 2 & 0.01499 & 0.02894 & $17_{10}$ & 0.363937 & (1, 3, 2, 2, 7, 2) & 2 & 0.01043 & 0.01635 \\
$12_{1}$ & 0.193438 & (3, 3, 6) & - & - & - & $18_{1}$ & 0.252412 & (1, 6, 11)  & - & - & - \\
$12_{2}$ & 0.198368 & (3, 6, 3)  & 1 & 0.00585 & 0.00585 & $18_{2}$ & 0.283157 & (1, 5, 12) & 1 & 0.00003 & 0.00003 \\
$12_{3}$ & 0.350225 & (4, 8)  & - & - & - & $18_{3}$ & 0.328686 & (1, 3, 2, 1, 4, 7)  & 1 & 0.01066 & 0.01066 \\
$12_{4}$ & 0.357584 & (2, 1, 1, 2, 6)  & 1 & 0.00877 & 0.00877 & $18_{4}$ & 0.351105 & (3, 3, 9, 3)  & - & - & - \\
$12_{5}$ & 0.373759 & (4, 4, 4)  & 1 & 0.00885 & 0.00885 & $18_{5}$ & 0.352151 & (6, 6, 6) & 1 & 0.00481 & 0.00481 \\
$12_{6}$ & 0.468791 & (2, 3, 7)  & 2 & 0.00873 & 0.01596 & $18_{6}$ & 0.356264 & (6, 12) & 2 & 0.00649 & 0.01005 \\
$12_{7}$ & 0.47031 & (2, 4, 4, 2)  & 2 & 0.00570 & 0.01122 & $18_{7}$ & 0.36608 & (1, 2, 1, 2, 1, 9, 2) & 1 & 0.01061 & 0.01061 \\
$12_{8}$ & 0.561931 & (2, 2, 2, 4, 2)  & 2 & 0.01887 & 0.03314 & $18_{8}$ & 0.3679 & (1, 3, 2, 2, 8, 2) & 2 & 0.00950 & 0.01319 \\
$12_{9}$ & 0.629558 & (1, 1, 2, 2, 6)  & 2 & 0.01636 & 0.02651 & $18_{9}$ & 0.477282 & (5, 13)  & - & - & - \\
$12_{10}$ & 0.635656 & (1, 2, 5, 2, 2)  & 2 & 0.02548 & 0.03253 & $18_{10}$ & 0.477282 & (5, 5, 8) & 1 & 0.00003 & 0.00003 \\
$13_{1}$ & 0.224324 & (2, 2, 9) & - & - & - & $19_{1}$ & 0.186265 & (1, 6, 6, 6)  & - & - & - \\
$13_{2}$ & 0.224324 & (4, 9)  & 1 & 0.00005 & 0.00005 & $19_{2}$ & 0.193174 & (1, 6, 12) & 1 & 0.00393 & 0.00393 \\
$13_{3}$ & 0.246359 & (3, 3, 2, 5) & - & - & - & $19_{3}$ & 0.33323 & (1, 4, 3, 11)  & - & - & - \\
$13_{4}$ & 0.246359 & (3, 4, 6)  & 1 & 0.00009 & 0.00009 & $19_{4}$ & 0.33323 & (1, 5, 2, 11) & 1 & 0.00012 & 0.00012 \\
$13_{5}$ & 0.297916 & (2, 1, 1, 4, 2, 3)  & 1 & 0.01209 & 0.01209 & $19_{5}$ & 0.346853 & (1, 4, 2, 1, 11) & 1 & 0.00909 & 0.00909 \\
$13_{6}$ & 0.550225 & (1, 1, 1, 2, 3, 4, 1)  & 2 & 0.01549 & 0.02795 & $19_{6}$ & 0.380323 & (1, 5, 4, 9) & 1 & 0.00007 & 0.00007 \\
$13_{7}$ & 0.553063 & (1, 3, 3, 6)  & 2 & 0.00928 & 0.01855 & $19_{7}$ & 0.386247 & (1, 5, 1, 4, 4, 4) & 1 & 0.00960 & 0.00960 \\
$13_{8}$ & 0.565349 & (3, 2, 8)  & 2 & 0.01182 & 0.02009 & $19_{8}$ & 0.438618 & (1, 2, 3, 2, 2, 5, 4) & 2 & 0.01325 & 0.02339 \\
$13_{9}$ & 0.569959 & (1, 2, 2, 2, 6)  & 3 & 0.01235 & 0.02504 & $19_{9}$ & 0.443897 & (1, 5, 2, 3, 4, 4) & 2 & 0.01230 & 0.02217 \\
$13_{10}$ & 0.583846 & (2, 1, 2, 4, 2, 2)  & 2 & 0.01809 & 0.03158 & $19_{10}$ & 0.44448 & (1, 4, 1, 2, 1, 6, 4) & 2 & 0.01332 & 0.02396 \\
$14_{1}$ & 0.17789 & (4, 4, 6) & - & - & - & $20_{1}$ & 0.219554 & (1, 6, 13)  & - & - & - \\
$14_{2}$ & 0.177914 & (4, 2, 4, 4)  & 1 & 0.00068 & 0.00068 & $20_{2}$ & 0.219596 & (1, 7, 12)  & - & - & - \\
$14_{3}$ & 0.367133 & (5, 9)  & - & - & - & $20_{3}$ & 0.295775 & (1, 2, 4, 1, 10, 2)  & 1 & 0.01144 & 0.01144 \\
$14_{4}$ & 0.367134 & (5, 9)  & 1 & 0.00011 & 0.00011 & $20_{4}$ & 0.429367 & (2, 2, 4, 2, 6, 4) & 1 & 0.01182 & 0.01182 \\
$14_{5}$ & 0.369135 & (2, 2, 1, 9)  & 1 & 0.00598 & 0.00598 & $20_{5}$ & 0.436846 & (2, 2, 4, 8, 4) & 2 & 0.00757 & 0.01127 \\
$14_{6}$ & 0.409639 & (3, 2, 2, 7)  & 1 & 0.00399 & 0.00399 & $20_{6}$ & 0.450123 & (2, 4, 2, 12) & 1 & 0.00786 & 0.00786 \\
$14_{7}$ & 0.411656 & (1, 2, 2, 3, 2, 4)  & 2 & 0.01291 & 0.01830 & $20_{7}$ & 0.450127 & (1, 1, 4, 1, 1, 12) & 2 & 0.00788 & 0.00907 \\
$14_{8}$ & 0.450731 & (2, 1, 2, 2, 2, 2, 2, 1)  & 2 & 0.01462 & 0.02122 & $20_{8}$ & 0.450553 & (2, 2, 2, 2, 2, 2, 8) & 1 & 0.01240 & 0.01240 \\
$14_{9}$ & 0.457754 & (1, 2, 2, 3, 2, 4)  & 2 & 0.01483 & 0.02498 & $20_{9}$ & 0.456025 & (1, 1, 1, 2, 1, 1, 1, 12) & 2 & 0.01075 & 0.01584 \\
$14_{10}$ & 0.530024 & (1, 2, 2, 1, 2, 6)  & 2 & 0.00883 & 0.01308 & $20_{10}$ & 0.468974 & (2, 4, 2, 6, 4, 2) & 3 & 0.01321 & 0.02834 \\
$15_{1}$ & 0.234126 & (4, 11)  & 1 & 0.00002 & 0.00002 & $30_{1}$ & 0.276668 & (2, 2, 4, 6, 16)  & - & - & - \\
$15_{2}$ & 0.247105 & (5, 10)  & - & - & - & $30_{2}$ & 0.276848 & (4, 2, 4, 4, 16)  & 1 & 0.00127 & 0.00127 \\
$15_{3}$ & 0.247735 & (5, 5, 5)  & 1 & 0.00153 & 0.00153 & $30_{3}$ & 0.278058 & (4, 2, 4, 4, 16) & 1 & 0.00173 & 0.00173 \\
$15_{4}$ & 0.301582 & (2, 2, 1, 2, 6, 2)  & 1 & 0.01087 & 0.01087 & $30_{4}$ & 0.278137 & (4, 4, 4, 2, 16) & 2 & 0.00168 & 0.00248 \\
$15_{5}$ & 0.429823 & (1, 1, 2, 1, 2, 4, 4)  & 1 & 0.00907 & 0.00907 & $30_{5}$ & 0.40736 & (2, 2, 7, 2, 17)  & - & - & - \\
$15_{6}$ & 0.429865 & (1, 1, 2, 1, 2, 4, 4)  & 2 & 0.00927 & 0.01243 & $30_{6}$ & 0.407375 & (1, 3, 2, 4, 3, 17) & 1 & 0.00051 & 0.00051 \\
$15_{7}$ & 0.432932 & (1, 4, 2, 4, 4)  & 1 & 0.00794 & 0.00794 & $30_{7}$ & 0.409361 & (4, 4, 2, 3, 17) & 1 & 0.00256 & 0.00256 \\
$15_{8}$ & 0.435103 & (1, 2, 2, 4, 4, 2)  & 2 & 0.00698 & 0.01148 & $30_{8}$ & 0.409366 & (2, 2, 3, 2, 4, 17) & 2 & 0.00257 & 0.00287 \\
$15_{9}$ & 0.445068 & (1, 2, 2, 4, 2, 2, 2)  & 2 & 0.01241 & 0.02152 & $30_{9}$ & 0.410704 & (2, 1, 1, 2, 4, 4, 1, 15) & 1 & 0.00824 & 0.00824 \\
$15_{10}$ & 0.496522 & (3, 3, 3, 6)  & 2 & 0.01096 & 0.02191 & $30_{10}$ & 0.418122 & (3, 1, 2, 6, 1, 1, 16) & 1 & 0.00930 & 0.00930 \\

  \end{tabular}
  \end{ruledtabular}
\end{table*}

\section{Further formulation details}
\subsection{Pullback}
\label{app:pullback}
The method of moving frames, or the pullback method, is a robust way to reduce a continuous symmetry in a dynamical system \cite{SIMINOS2011187}, which was formalised by Cartan \cite{cartan1935}. The pullback operator for states $\mathbf x \in \mathcal M$
\begin{equation}
    \mathcal P: \mathcal M \to \Bar{\mathcal{M}},
\end{equation}
sends every point on our manifold of possible point vortex configurations $\mathcal{M}$ to a representative point on a slice of the manifold $\Bar{\mathcal{M}} \subset \mathcal{M}$. 
The $N$ vortex system has a $2N$ dimensional manifold and a continuous symmetry group $G$ (rotation). 
This slice is then a $2N-1$ dimensional submanifold which intersects all the group orbits of $G$ of $\mathbf{x}$ transversally and at most once. 
The slice is defined as the hyperplane normal to the tangent of $G$ at the slice fixing point $\Bar{\mathbf{x}}'$:
\begin{equation}
    (\Bar{\mathbf{x}} - \Bar{\mathbf{x}}') \cdot \boldsymbol \tau' = 0.
    \label{eq:general_slice_condition}
\end{equation}
The group tangent at $\Bar{\mathbf{x}}'$ is defined as $\boldsymbol{\tau}' = \mathbf T\Bar{\mathbf{x}}'$, where $\mathbf T$ is the generator of infinitesimal rotations for our system, such that $\mathbf G(\theta) = \exp{(\theta \mathbf T)}$ is an element of $G$. 
$\mathbf T$ is a $2N\times 2N$ matrix as:
\begin{equation*}
    \mathbf T = \begin{bmatrix}
    0 & -1 & 0 & 0 & \dots & 0 & 0 \\
    1 & 0 & 0 & 0 & \dots & 0 & 0 \\
    0 & 0 & 0 & -1 & 0 & \dots & 0 \\
    0 & 0 & 1 & 0 & 0 & \dots & 0 \\
    & & & \vdots \\
    0 & & & \dots & & 0 & -1 \\
    0 & & & \dots & & 1 & 0 \\
    \end{bmatrix},
\end{equation*}
and the state vector $\mathbf{x}$ is a 2$N$ dimensional vector $( x_1, y_1, x_2, y_2, \dots, x_N, y_N )$. 
Expanding the exponential of the above generator, one gets the rotation matrix which rotates each vortex $( x_i, y_i )$ counter-clockwise by $\theta$: 
\begin{equation*}
    \mathbf G(\theta) = \begin{bmatrix}
    \cos{\theta} & -\sin{\theta} & 0 & 0 & \dots & 0 & 0 \\
    \sin{\theta} & \cos{\theta} & 0 & 0 & \dots & 0 & 0 \\
    0 & 0 & \cos{\theta} & -\sin{\theta} & 0 & \dots & 0 \\
    0 & 0 & \sin{\theta} & \cos{\theta} & 0 & \dots & 0 \\
    & & & \vdots \\
    0 & & & \dots & & \cos{\theta} & -\sin{\theta} \\
    0 & & & \dots & & \sin{\theta} & \cos{\theta} \\
    \end{bmatrix}.
\end{equation*}
The pullback operator $\mathcal P$ is defined by a group element $\mathbf G(\theta)$, and it remains to find an appropriate expression for $\theta(\mathbf x)$.
The slice condition in equation (\ref{eq:general_slice_condition}) can be simplified to
\begin{equation}
    (\mathbf G(\theta)\mathbf{x})^T \mathbf T\Bar{\mathbf{x}}' = 0,
    \label{eq:simple_slice_condition}
\end{equation}
as $\mathbf T^T = -\mathbf T$ is antisymmetric so that $\Bar{\mathbf{x}}'^T  \mathbf T\Bar{\mathbf{x}}' = 0$.
We now use equation (\ref{eq:simple_slice_condition}) to find an expression for $\theta(\mathbf{x})$. 

Without loss of generality we pick the slice fixing point $\Bar{\mathbf{x}}' = ( 0, \Bar{y}'_1, \Bar{x}'_2, \Bar{y}'_2, \dots, \Bar{x}'_N, \Bar{y}'_N )$, 
and in practice, the simplest slice fixing point is $\Bar{\mathbf{x}}' = ( 0, 1, 0,0, \dots, 0,0 ) $. 
Then the expression for $\theta$ follows from (\ref{eq:simple_slice_condition}) as $\arctan{(x_1/y_1)}$ (respecting the quadrants in the $(x_1, y_1)$ plane). This pullback operation rotates a system so that the vortex at index 1 has the same $x$-component as the centre of vorticity of the system.

\subsection{DNEB method}
\label{app:dneb}
Gradients of the loss in (\ref{eqn:vortex_final_chain_loss}) are `nudged' to counteract two common issues: `sliding-down' and `corner-cutting'. `Sliding-down' occurs when the intermediate points move towards the minima. The nudging approach counteracts this by only considering the portion of the gradient of $V$ orthogonal to the direction of the pathway. `Corner-cutting' occurs when the path is highly curved, and the spring potential causes the intermediate states to cut out this highly curved bend. The nudging approach counteracts this by only considering the portion of the gradient of $\Tilde{V}$ parallel to the direction of the pathway. The discussion here follows that given in Trygubenko and Wales \cite{trygubenko2004} and the supplementary information for Goodrich et al \cite{goodrich2021}.

Formalising this, let $\hat{\bm{\tau}}_i$ be the unit tangent vector to the chain at $\mathbf{x}_i$:
\begin{widetext}
\begin{equation}
    \hat{\bm{\tau}}_i = \begin{cases}
        \dfrac{(j-i)(\mathbf{x}_j - \mathbf{x}_i)}{\| \mathbf{x}_j - \mathbf{x}_i \|}   \qquad  \text{If exactly one of the neighbours $j$ satisfies } \mathcal{F}(\mathbf{x}_j) >  \mathcal{F}(\mathbf{x}_i) \\
        \dfrac{\mathbf{x}_{i+1} - \mathbf{x}_{i-1}}{ \| \mathbf{x}_{i+1} - \mathbf{x}_{i-1} \|}  \qquad \text{Otherwise (i.e. is a local extremum)}
    \end{cases}
\end{equation}
\end{widetext}
Let $\bm{\nabla}_i$ be the gradient with respect to the state vector of $\mathbf{x}_i$. 
\begin{align}
    \bm{g}_i &= \bm{\nabla}_i V = \bm{g}_i^{\|} + \bm{g}_i^{\perp} \\
    \Tilde{\bm{g}}_i &= \bm{\nabla}_i \Tilde{V} = \Tilde{\bm{g}}_i^{\|} + \Tilde{\bm{g}}_i^{\perp}
\end{align}
where
\begin{align}
    \bm{v}^{\|} &= \left( \bm{v} \cdot \hat{\bm{\tau}}_i\right) \hat{\bm{\tau}}_i \\
    \bm{v}^{\perp} &= \bm{v} - \bm{v}^{\|}
\end{align}
The nudged elastic band (NEB) method uses 
\begin{equation}
    \bm{g}_i = \bm{g}_i^{\perp} + \Tilde{\bm{g}}_i^{\|}
\end{equation}
to optimise $\mathcal{L}_P$. The doubly-nudged elastic band (DNEB) method only projects out some of the $\Tilde{\bm{g}}_i^{\perp}$ term so that 
\begin{equation}
    \bm{g}_i = \bm{g}_i^{\perp} + \Tilde{\bm{g}}_i^{\|} + (\Tilde{\bm{g}}_i^{\perp} - (\Tilde{\bm{g}}_i^{\perp} \cdot \hat{\bm{g}}_i^{\perp}) \hat{\bm{g}}_i^{\perp} )
\end{equation}
is instead used to optimise $\mathcal{L}_P$.

\section*{Data Availability Statement}

The data that support the findings of
this study are available from the
corresponding author upon reasonable
request.

\begin{acknowledgments}

This research has been supported by the UK Engineering and Physical Sciences Research Council through the MAC-MIGS Centre for Doctoral Training (EP/S023291/1). Computations were performed on  the Cirrus UK National Tier-2 HPC Service at EPCC (http://www.cirrus.ac.uk). 
JP acknowledges support from UKRI via grant EP/Y004094/1.

\end{acknowledgments}

\bibliography{bib.bib}

\end{document}